\documentclass[pdflatex,sn-mathphys-num,iicol]{sn-jnl}% Math and Physical Sciences Numbered Reference Style 
\usepackage{graphicx}
\usepackage{multirow}
\usepackage{amsmath,amssymb,amsfonts}
\usepackage{amsthm}
\usepackage{mathrsfs}
\usepackage[title]{appendix}
\usepackage{xcolor}
\usepackage{textcomp}
\usepackage{manyfoot}
\usepackage{booktabs}
\usepackage{algorithm}
\usepackage{algorithmicx}
\usepackage{algpseudocode}
\usepackage{listings}
\usepackage{subfigure}
\usepackage{cleveref}
\usepackage{tcolorbox}
\usepackage{hyperref}
\usepackage{lmodern}
\usepackage{anyfontsize}

\newtcolorbox{rqbox}[1][]{colback=blue!5!white, colframe=blue!75!black, fonttitle=\bfseries, coltitle=black, boxrule=0.5mm, arc=4mm, auto outer arc, width=\linewidth, #1}

\begin{document}

\title[Reformulating Regression Test Suite Optimization using Quantum Annealing - an Empirical Study]{Reformulating Regression Test Suite Optimization using Quantum Annealing - an Empirical Study}

%%=============================================================%%
%% GivenName	-> \fnm{Joergen W.}
%% Particle	-> \spfx{van der} -> surname prefix
%% FamilyName	-> \sur{Ploeg}
%% Suffix	-> \sfx{IV}
%% \author*[1,2]{\fnm{Joergen W.} \spfx{van der} \sur{Ploeg} 
%%  \sfx{IV}}\email{iauthor@gmail.com}
%%=============================================================%%

\author*[1]{\fnm{Antonio} \sur{Trovato}}\email{atrovato@unisa.it}

\author[1]{\fnm{Manuel} \sur{De Stefano}}\email{madestefano@unisa.it}

\author[1,2]{\fnm{Fabiano} \sur{Pecorelli}}\email{fabiano.pecorelli@unipegaso.it}

\author[1]{\fnm{Dario} \sur{Di Nucci}}\email{ddinucci@unisa.it}

\author[1]{\fnm{Andrea} \sur{De Lucia}}\email{adelucia@unisa.it}

\affil[1]{\orgdiv{SeSa Lab}, \orgname{University of Salerno}, \orgaddress{\city{Salerno}, \state{Italy}}}

\affil[2]{\orgdiv{Dep. of Information Science and Technology}, \orgname{Pegaso Digital University}, \orgaddress{\city{Naples}, \state{Italy}}}

%%==================================%%
%% Sample for unstructured abstract %%
%%==================================%%

\abstract{Maintaining software quality is crucial in the dynamic landscape of software development. Regression testing ensures that software works as expected after changes are implemented. However, re-executing all test cases for every modification is often impractical and costly, particularly for large systems.

Although very effective, traditional \textit{test suite optimization} techniques are often impractical in resource-constrained scenarios, as they are computationally expensive. Hence, \textit{quantum computing} solutions have been developed to improve their efficiency but have shown drawbacks in terms of effectiveness.

We propose reformulating the regression test case selection problem to use quantum computation techniques better. Our objectives are (i) to provide more efficient solutions than traditional methods and (ii) to improve the effectiveness of previously proposed quantum-based solutions.

We propose \textit{SelectQA}, a \textit{quantum annealing} approach that can outperform the quantum-based approach \textit{BootQA} in terms of effectiveness while obtaining results comparable to those of the classic \textit{Additional Greedy} and \textit{DIV-GA} approaches. Regarding efficiency, SelectQA outperforms DIV-GA and has similar results with the Additional Greedy algorithm but is exceeded by BootQA.}


%%\pacs[JEL Classification]{D8, H51}

%%\pacs[MSC Classification]{35A01, 65L10, 65L12, 65L20, 65L70}

\maketitle

\section{Introduction}
\label{sec:intro}

In the ever-evolving landscape of software development, software quality assurance is of fundamental importance. As software systems undergo continuous modifications and enhancements, ensuring these changes do not introduce unintended side effects or defects becomes imperative. In response to this need, regression testing~\cite{yoo2012regression} verifies whether previously developed and tested software components still work as expected after a change is performed. 

Ideal regression testing would re-run all the available test cases of a given software system. However, in addition to being potentially very costly, this could even be impractical in some cases~\cite{rothermel1998empirical,perrouin2012pairwise}, especially for large systems and when the re-execution of entire test suites occurs for every single modification. Several approaches have been suggested to streamline the regression testing process, such as selecting a potentially minimal subset of test cases from the test suite based on specific testing criteria~\cite{yoo2007pareto,yoo2010using,yoo2011highly,wang2013minimizing,panichella2014improving,arrieta2019pareto,xue2020multi}. Alternatively, strategies include prioritizing the execution of test cases, aiming first to run those anticipated to uncover faults earlier in the testing cycle~\cite{li2007search,assunccao2014multi,epitropakis2015empirical,di2018test}.

In this scenario, \textit{test case selection}~\cite{engstrom2010systematic,yoo2012regression} is one of the most widely investigated \textit{test suite optimization} approaches. It consists of choosing a subset of test cases from a pool of possibilities (i.e., the test suite), ensuring comprehensive coverage and efficient use of testing resources.

While the significance of test case selection is undeniable, traditional techniques face serious challenges, primarily related to computing costs. Conventional approaches, often rooted in optimization algorithms, such as greedy algorithms~\cite{yoo2007pareto} and search-based approaches like DIV-GA~\cite{panichella2014improving}, demand extensive computational resources to deliver optimal results. The computational burden reduces the efficiency of these methods and poses practical limitations, especially in scenarios where resource-intensive testing is unavailable.

Relying on quantum computing could overcome the limitations of traditional techniques. This new computation technology harnesses the principles of quantum mechanics to process information in ways fundamentally different from classical computers. The intrinsic parallelism and exponential computational capacity of quantum systems offer a potential breakthrough for overcoming the resource constraints associated with traditional test case selection techniques~\cite{hoare2005grand,knight2018serious}. 

The current state-of-the-art quantum computing techniques for test suite optimization are represented by \emph{BootQA}~\cite{bootqa} and IGDec-QAOA~\cite{qaoa_tco}. These techniques represent the first quantum approaches developed for the test suite optimization problem. BootQA resolves the problem through \textit{quantum annealing}, whereas IGDec-QAOA uses the quantum approximate optimization algorithm (QAOA). We focus on quantum annealing as a first step. Developing QAOA strategies is way more challenging due to the significant limitations of current gate-based quantum computers compared to annealing ones. We will perform such a comparison in future work.

This paper proposes SelectQA, a novel quantum computing approach that solves the Test Case Selection (TCS) problem. Like for \emph{BootQA}, the algorithm is implemented through the D-Wave environment~\cite{dwave_env}, which can solve NP-hard combinatorial problems through ``\textit{Quantum Simulated Annealing}''~\cite{farhi2014quantum, Das2008Colloquium}.

Leveraging Quantum Simulated Annealing~\cite{kadowaki1998quantum,suzuki2009comparison} and executing different runs of experimentation, SelectQA can produce great (and statistically reliable) results, both in terms of effectiveness and efficiency. In particular, in terms of effectiveness, SelectQA outperforms BootQA while producing results comparable to those of additional greedy and DIV-GA. In terms of efficiency, SelectQA has shown a practically constant total execution time, regardless of the problem size, demonstrating its great scalability. In this perspective, SelectQA outperforms DIV-GA but is outperformed by BootQA. Additional greedy performs faster than SelectQA in two out of four cases, showing its remarkable ability to solve small problems while impractical for larger ones. 

These results promise to address the computational challenges plaguing traditional methods. As quantum computing continues to emerge as a driving force in optimization tasks, this work has aimed to provide a significant stride toward advancing the efficiency and effectiveness of regression testing in contemporary software development paradigms, taking additional steps into quantum software engineering research.

\textbf{Overview of the paper:} \Cref{sec:bg} introduces with much more detail the problems of test case selection that SelectQA aims to solve and the methods that represent the current state-of-the-art with which to compare SelectQA; \Cref{sec:toolname} describes the actual mathematical formulations behind the work of SelectQA; \Cref{emp_eval} first introduces the goals, metrics, and configurations planned for the experiments; then, it shows the actual results with the correspondent conclusions. \Cref{sec6:ttv} makes a deeper analysis of all the threats to the validity of this study; finally, \Cref{sec:conclusion} concludes the paper and envisions future work.

\section{Background}
\label{sec:bg}
This section introduces test case selection, its multi-objective formulation, and the usage of classical and quantum optimizations to solve it.

\subsection{Test Case Selection as Optimization Problem}
\label{sec2.1:tcs}
Test case selection focuses on selecting a subset of a test suite to test software changes, i.e., to test whether unmodified parts of a program continue to work correctly after changes involving other parts~\cite{rothermel1996analyzing}. 
Various techniques, such as Integer Programming~\cite{fischer1977test}, symbolic execution~\cite{yau1987method}, data flow analysis~\cite{rothermel1993safe}, dependence graph-based methods~\cite{bates1993incremental}, and flow graph-based approaches~\cite{rothermel1996analyzing}, can be employed to identify the modified portions of the software. Once test cases covering the unchanged program segments are pinpointed using a specific technique, an optimization strategy can \textit{select a minimal set of these test cases based on certain testing criteria} (e.g., branch coverage). The ultimate aim is to reduce the costs associated with regression testing.

While initially test case selection was considered a single-objective optimization problem, advancements in the method revealed that optimizing only one objective is insufficient, as many tests often need to satisfy multiple criteria simultaneously. Therefore, Yoo et al.~\cite{yoo2007pareto} introduced the idea of using Pareto sets to address a multi-objective version of the test case selection problem.
In particular, in their multi-objective variant, the goal is to build a \textit{Pareto-efficient} subset of the starting test suite, leveraging different selection criteria, given a set of components to test. 

Let  $\Gamma = \{\tau_1,\tau_2,...,\tau_n\}$ be the starting test suite and let $F = \{f_1,f_2,...,f_m\}$ the set of objective functions that mathematically formalize the criteria for the selection process. The multi-objective formulation of the test case selection problem can be seen as selecting a subset $\Gamma' \subseteq \Gamma$ such that $\Gamma'$ is the Pareto-optimal set concerning the objective functions in $F$.

What emerges from the definition above is that the optimality of a solution is measured using the concepts of Pareto optimality and dominance. A solution $X$ is said to be \textit{Pareto-optimal} if and only if it is \textit{non-dominated} by any other solution in the entire search space, which means that any other solution \textit{Y} that improves one of the objective functions worsens other objectives. The set of all the non-dominated solutions is called the \textit{Pareto-optimal set}, and the corresponding \textit{objective vectors} form the \textit{Pareto frontier}.

Multi-objective genetic algorithms, namely MOGAs, are often applied to solve this problem. Yoo and Harman~\cite{yoo2007pareto,yoo2010using} explore test case selection through an experimental study, employing the additional greedy algorithm, NSGA-II algorithm, and a variant of NSGA-II, called the vNSGA-II algorithm. The findings indicate that the additional greedy algorithm is more suitable for single-objective test case selection. In contrast, NSGA-II and vNSGA-II algorithms perform better in multi-objective test case selection.

They~\cite{yoo2007pareto,yoo2010using} also considered three contrasting criteria for the selection: code coverage, execution time, and fault history information. Then, empirical results showed no clear winner between MOGAs and additional greedy~\cite{yoo2007pareto}, nor did a hybrid approach produce better results~\cite{yoo2010using}.

Later on, Panichella et al.~\cite{panichella2014improving} introduced DIV-GA, an algorithm that improves the NSGA-II algorithm by injecting diversity into the genetic algorithm, reducing the genetic drift phenomenon in NSGA-II and enhancing the effectiveness and efficiency of test case selection.

\subsection{Quantum Optimization for Test Case Selection}
%\subsubsection{Quantum Optimization and Quantum Annealing}
The ``quantum era'' promises to impact program computation so that even NP-hard problems will be solvable.

%As an example, the programming framework of IBM aims to allow developers to develop, design, and execute quantum software on cloud-based quantum systems; off-the-shelf instruments, APIs, programming languages, and development toolkits have emerged in the last years to support the growing quantum developing community.
%Quantum computing is a growing and prolific research topic, as it promises to solve very complex computational problems in ways much more efficient than classical machines. In theory, quantum machines would be able to outperform even supercomputers, achieving results that are definitely out of their reach. However, due to the limitations of currently available quantum computers, it is impossible to officially demonstrate quantum computing superiority.

Qubits, generally represented by the electron spin, can represent different variable values at the same time. While a bit can represent a zero or a one, a qubit can be in a state of either of the two classical values with a certain probability. The actual value of a qubit is then known with the process of measurement, where the qubit is collapsed to a classical bit; when this happens, the qubit must be reset to make it reusable as a qubit. 

Quantum optimization algorithms, such as Grover's Algorithm~\cite{grover1996fast}, use quantum oracles to search for unknown spaces with linear complexity. Quantum environments, e.g., D-Wave Quantum Leap, optimize NP-hard problems through adiabatic quantum optimization, defining the problem as a quantum system and its energy as a Hamiltonian to find the lowest energy solution.

Quantum approximate optimization algorithms (QAOAs) showed great potential in solving optimization algorithms, exploiting the capabilities of both classical and quantum computations. Wang et al.~\cite{qaoa_tco} proposed IGDec-QAOA, a method that resolves the test suite optimization problem as a QAOA problem, reaching significant effectivity levels; as explained earlier, we decided to leave further comparisons with IGDec-QAOA as future work.

Resolving optimization problems is the main focus of quantum annealers, a special kind of quantum computer designed for this specific goal, which applies the quantum version of simulated annealing: \textit{Quantum Annealing} (QA). While the gate model divides problems into a sequence of operations on the qubits, quantum annealing translates the problem into a quantum system of qubits to find the minimum energy configuration through a gradual quantum state transition.

The method takes the steps from the \textit{Adiabatic Theorem~\cite{at1}\cite{at2}, according to which, starting a quantum system from its ground state, which represents its minimum energy state, if the \textit{Hamiltonian} of the system changes slowly enough, the system will remain in its ground state during the whole evolution process.
It solves combinatorial optimization problems by starting from a trivial- or easy-to-compute ground state related to an initial Hamiltonian and then slowly evolving the system towards a final Hamiltonian representing the problem to solve; thus, its ground state would represent the solution to the problem.}

%\Cref{fig:annealing} depicts the quantum annealing process. 

Quantum annealing formalizes combinatorial optimization problems as single-objective ones.
In particular, D-Wave uses the QUBO (Quadratic Unconstrained Binary Optimization) model to represent the objective function to minimize (i.e., the final Hamiltonian).
It allows evaluating multiple states simultaneously and finding optimal solutions more efficiently, especially in complex landscapes where classical methods may struggle~\cite{suzuki2009comparison}. This heuristic algorithm cannot guarantee finding the optimal solution; therefore, it uses multiple sampling processes to generate candidate solutions in a single execution.

The main reason behind the choice of quantum annealing instead of gate-based quantum algorithms is the limitations of the former in available qubits, which are insufficient to resolve complex optimization problems. 
In particular, the two quantum annealing approaches experimented with in this work, BootQA and SelectQA, use different strategies to cope with highly complex and qubit-demanding problems.
While BootQA performs a custom local decomposition strategy, that is, bootstrapping sampling, to obtain subproblems that can be solved directly by the QPU of the D-Wave Advanced System~\cite{adv_sys}, SelectQA relies on the D-Wave Hybrid Solver Service (HSS)~\cite{tech_rep}.
HSS takes the original QUBO model $Q$ of the problem to solve as input and runs different hybrid solvers running on both CPUs and QPUs in parallel. The hybrid solvers query the D-Wave's Advantage QPU with sub-instances of the initial $Q$ using Query Modules (QMs). The hybrid solvers interpret the results as suggestions about promising areas of the search space.
%It involves a journey through an energy landscape, starting from a broad valley representing a superposition state and gradually shifting to a double-well potential state. This transition signifies the convergence to the lowest energy solution, representing the optimal solution. Quantum annealing's ability to explore a vast array of potential solutions efficiently improves the efficiency of solving complex optimization problems beyond classical computing's scope.

%\begin{figure*}[t!]
%    \centering
%    \includegraphics[width=0.7\textwidth]{figures/annealing.png}
%    \caption{Quantum annealing as reported by D-Wave~\cite{annealing_process}}
%    \label{fig:annealing}
%\end{figure*}

%Before introducing BootQA and the SelectQA proposed in this work, a brief introduction to QUBO problems is needed.

Wang et al.~\cite{bootqa} pioneered using quantum annealing for test suite optimization, proposing  BootQA.
This quantum annealing tool is designed to resolve the test suite optimization problem using quantum annealing and circumvent the physical limitations of currently available quantum annealers. It is an alternative approach whose first purpose is \textit{minimizing the number of test cases, promoting at the same time other defined objectives}. BootQA answers two research challenges: (i) designing a brand new QUBO model for the test suite optimization problem and (ii) designing a new method of qubit optimization.
Due to their limited number of available qubits, the currently available quantum annealers can resolve large-scale problems only relying on classical computations as hybrid solvers do, like the one used by SelectQA~\cite{tech_rep}.

\begin{figure}[t!]
    \centering
    \includegraphics[width=0.50\columnwidth]{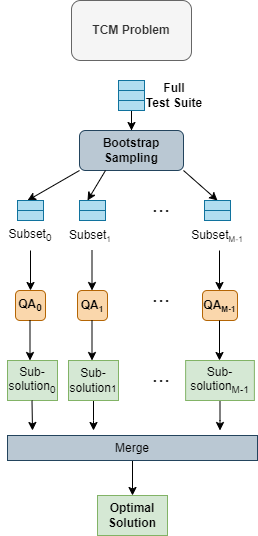}
    \caption{BootQA~\cite{bootqa} overview}
    \label{fig:bootqa}
\end{figure}

Therefore, BootQA applies \textit{bootstrap sampling} to decompose the original problem into smaller sub-problems (see \Cref{fig:bootqa}).
The idea is to build small sub-problems by sampling subsets of test cases from the original test suite, ensuring that each generated sub-problem is solvable using the available quantum annealers.

Starting from the original test suite $TS$, BootQA decomposes it into $m$ smaller test suites of size $n$. Of course, $n$ is chosen based on the limitations of QA hardware. The $m$ subsets are sampled \textit{randomly}, meaning the same test case can appear in different sampled subsets. So, to make BootQA cover an exhaustive percentage of the test cases (depicted by the hyperparameter $\beta$), $m$ is empirically set large enough. The resulting $m$ subsets are used as input for the sub-problem QUBO formulation, then individually resolved by the annealing process, resulting in a sub-solution.
For each sub-problem, an overall objective, which is an instance of the generic QUBO function, is constructed. Eventually, the different $m$ sub-solutions derived from each sub-problem are merged into one single solution. In particular, each selected test case in each sub-problem is considered selected in the final solution.

Configuring BootQA to achieve the best subset of test cases could be challenging; therefore, we contacted the authors of the original paper~\cite{bootqa} to ensure a correct configuration. Although the $n$ parameter can be easily chosen, configuring the corresponding optimal $m$ parameter is way more tedious. The configuration should assume values that allow the sampling process to obtain adequate coverage compared to the original test suite. The sampling should be conducted several times with different ${(m,n)}$ pairs to select the optimal value, spending large amounts of execution time and possibly exceeding the time supplied by D-Wave. Furthermore, the random nature of sampling hinders direct control of the test case coverage, leading to the sampled sub-suites not representing the initial test suites and unstable solutions that do not represent good trade-offs between the objectives.

\begin{table}[ht]
    \centering
    \caption{Algorithm definitions}
    \begin{tabular}{p{0.09\linewidth} p{0.800\linewidth} p{0.1\linewidth}}
        \toprule
        \textbf{\textcolor{black}{\#}} & \textbf{\textcolor{black}{Definition}}\\
        \bottomrule
        $i$ & unique index identifying a test case\\
        $k$ & unique index representing a statement\\
        $\Gamma$ & the starting test suite \\
        $T_k$ & list of all test cases running the $k$-th statement \\
        $x_i$ & binary variable that specifies if the $i$-th test case is part of the solution\\
        $cost(\tau_i)$ & normalized execution cost of the $i$-th test\\
        $e_i$ & binary variable that indicates whether the $i$-th test case has detected errors in the past\\
        $f_i$ & the failure rate, i.e., the percentage of times a test case spotted a failure in the past\\
        $\alpha$ & weight factor to modulate the contrast ratio between the objectives of the problem\\
        $P$ & penalty coefficient to regulate the weight of the constraints of the problem\\
        \bottomrule
    \end{tabular}
    \label{table:constraints}
\end{table}

\section{SelectQA}
\label{sec:toolname}

This section introduces SelectQA, a method based on quantum annealing to solve the test case selection problem, as described in \Cref{sec2.1:tcs}.
To fully understand how the problem has been modeled, we provide the definitions of variables and functions necessary for its formulation in \Cref{table:constraints}. 

\subsection{General QUBO Formulation}
\label{algorithm_approach}
Formulating the test case selection problem as a QUBO problem involves a series of steps seamlessly integrating into a coherent framework. Initially, the primary objectives are set out, focusing on minimizing the execution cost of the test suite while maximizing its efficacy in detecting failures. We aim to compare the SelectQA approach with classical and quantum approaches, so we developed two different formulations: one coherent with the objectives chosen by Panichella et al.~\cite{panichella2014improving}, the other coherent with the objectives chosen by Wang et al.~\cite{bootqa}.
In the three-objective version coherent with Panichella et al.~\cite{panichella2014improving}, each test case is characterized by its execution cost, historical information about its ability to detect failures, and the statements it covers.
In the two-objective version coherent with Wange et al.~\cite{bootqa}, each test case is characterized by its execution cost and the failure rate, i.e., the \textit{percentage} of failures spotted by a test case based on the fault history.
The two versions implement an algorithm that resolves the \textit{Minimum Vertex Cover} problem. Furthermore, integrating the statement coverage constraint into the linear equation aligns with Serrano et al.~\cite{marisma_qa}.

\subsubsection{The Three-Objective Version of SelectQA}
The goal is to find the smallest subset of test cases that collectively covers all necessary program statements (those covered by the initial suite $\Gamma$), referred to as set $S$. Each test case is essentially a subset of $S$ (in terms of the statements it covers), and the challenge is to identify the minimal collection of these subsets that efficiently covers the entire set $S$.

In the QUBO framework, this problem is expressed using binary variables (0 or 1) to represent the inclusion or exclusion of a test case in the final suite. The QUBO problem is then described by a Hamiltonian function or a Binary Quadratic Model (BQM), which incorporates the linear impacts of each test case and the quadratic terms representing interactions between different test cases, such as overlapping coverage of program statements.
Following the methodologies proposed by Glover et al.~\cite{qubo_tutorial}, we transform the linear objectives of the test case selection problem into a quadratic form by creating a Hamiltonian that encapsulates the individual contributions of each test case and their interrelations.
The formulated Hamiltonian is processed using quantum annealing, specifically the D-Wave system, designed for solving QUBO problems. This approach allows efficient solution space exploration to find the optimal subset of test cases. This method leverages quantum computing capabilities and promises more efficient solutions than classical algorithms, particularly for extensive and complex test suites.

The first goal of SelectQA consists of minimizing the overall execution cost of the resulting (sub-) test suite. Considering $\tau_i$ a generic test case and $cost(\tau_i)$ its corresponding \textbf{normalized} execution cost, the function that represents the first goal to minimize is formalized as follows:

\begin{align}
\alpha\sum_{i=1}^{|\Gamma|} [x_i \cdot cost(\tau_i)]
\end{align}

The second objective of SelectQA consists of maximizing the overall fault coverage of the resulting (sub-) test suite. Here, having $e_i$ indicating whether the $i-th$ test did detected a failure in the past, the function that represents the second goal to maximize is formalized below as:

\begin{align}
(1-\alpha)\sum_{i=1}^{|\Gamma|} (e_i \cdot x_i)
\end{align}

Since we want to minimize the overall objective Hamiltonian, the second objective function is converted to the following one to minimize:

\begin{align}
-(1-\alpha)\sum_{i=1}^{|\Gamma|} (e_i \cdot x_i)
\end{align}

Please note the presence of the $\alpha$ coefficient within the two target functions.
This coefficient is a weight factor ($0 <  \alpha < 1$) to establish preference towards one goal rather than another using a weighted sum ~\cite{multi_opt}.
$ \alpha = 0.5$ means that two objectives are equally important, whereas other values represent the more relevance of an objective over the other.
The final function to be minimized consists of two parts and is formulated as follows:

\begin{align}
H = \alpha\sum_{i=1}^{|\Gamma|} [x_i \cdot cost(\tau_i)]-(1-\alpha)\sum_{i=1}^{|\Gamma|} (e_i \cdot x_i)
\end{align}

We introduce a critical constraint to maintain the coverage level of the initial test suite: each program statement executed in the original suite must be covered by at least one test case of the final selection.
This constraint is essential to ensure that the final suite still comprehensively covers all necessary program statements despite reducing the number of test cases.
This step allows maintaining the integrity and effectiveness of the test coverage by adding the following constraint:

\begin{align}
\sum_{i \in T_k} (x_i) \ge 1
\end{align}

The constraint states that at least one test case that executes the $k$ statement must be selected. 
The list of test cases that run the $k$-th statement is $T_k$.
Since a program has several statements, we apply the constraint for each statement to cover and obtain the following constraints:

\begin{align}
\sum_{k}(\sum_{i \in T_k} (x_i-1)^2)
\end{align}

Given the previously described objectives and constraints, we construct a Hamiltonian expression to transform the linear test case selection problem into a QUBO problem.
To this end, we add a penalty constant ($P$)~\cite{qubo_tutorial}, which balances the importance of constraints within the Hamiltonian. 
The use of penalties is crucial when dealing with problems that include additional constraints other than the one requiring variables to be binary. This kind of problem can be re-formulated as QUBOs leveraging a penalty coefficient in the objective function, which represents an alternative to the explicit use of separated constraints. For minimizing objective functions, penalties are equal to zero for acceptable solutions and equal to some positive value for unacceptable solutions. In other words, the optimizer itself, by the introduction of penalties, will search for the solution to avoid incurring the penalties.
Empirically, setting the $P$ penalty weight can be done using the Upper Bound strategy, which consists of setting $P$ to be slightly higher than the maximum possible value achievable by the objective function. This step ensures that the penalty aligns with the application domain and significantly influences the solution process \cite{penalty_weights}. Hence, we have:

\begin{align}
H =&\ \alpha\sum_{i=1}^{|\Gamma|} [x_i \cdot cost(\tau_i)] \notag \\
  & -(1-\alpha)\sum_{i=1}^{|\Gamma|} (e_i \cdot x_i) \notag \\
  & + P \cdot \sum_{k} \left(\sum_{i \in T_k} (x_i-1)^2\right)
\end{align}

The last part of the equation ($P \cdot \sum_{k} \left(\sum_{i \in T_k} (x_i-1)^2\right)$), representing the constraints can be simplified as follows:

\begin{align}
P \cdot \sum_{k}(\sum_{i,j \in T_k} (x_i^2 + 1^2 + 2 x_i x_j - 2 x_j))
\end{align}

% Removing squares and irrelevant constants:

% \vspace{2mm}
% \begin{align}
% P \cdot \sum_{k}(\sum_{i,j \in T_k} (-x_i + 2 x_i x_j))
% \end{align}
% \vspace{2mm}

which can be further simplified as follows:

\begin{align}
\sum_{k}(\sum_{i,j \in T_k} (-P x_i + 2 P x_i x_j))
\end{align}

The resulting BQM expression of the Hamiltonian (i.e., QUBO in this case) is the following:

\begin{align}
H =&\ \alpha\sum_{i=1}^{|\Gamma|} x_i \cdot cost(\tau_i) \notag \\
  & -(1-\alpha)\sum_{i=1}^{|\Gamma|} (e_i \cdot x_i) \notag \\
  & + \sum_{k}\sum_{i,j \in T_k} (-P x_i + 2 P x_i x_j)
\end{align}

\subsubsection{The Two-Objective Version of SelectQA}
\label{sec:3.2.2}
The procedure to obtain a two-objective formulation of the problem to compare it with the same dataset used by BootQA is the same. In this case, we have a two-objective linear problem where we want to minimize the final suite execution cost while maximizing its failure rate. Without repeating the previous procedure, we formulate the two objectives and combine them into a linear equation. 
We obtain:

\begin{align}
H &= \alpha\sum_{i=1}^{|\Gamma|} x_i \cdot cost(\tau_i)\\ \notag 
&-(1-\alpha)\sum_{i=1}^{|\Gamma|} (f_i \cdot x_i)
\end{align}

Where the costs are normalized, and $f_i$ is a percentage value representing the failure rate.

\section{Empirical Evaluation}
\label{emp_eval}
This section describes our goal, research questions, and research methods.

\subsection{Goal and Research Questions}
The \textit{goal} is to evaluate the efficiency and effectiveness of the quantum-annealing-based test case selection method SelectQA and compare it to classical and quantum state-of-the-art approaches. The perspective is of researchers and practitioners: while the former are interested in improving state-of-the-art and classical computing techniques, the latter are interested in having a practically exploitable solution to their testing problems. We aim to answer the following research questions:

{\centering
    \begin{rqbox}
        \textbf{RQ$_1$}: Is SelectQA more \textbf{effective} than traditional state-of-the-art methods and than the BootQA method?
    \end{rqbox}
}

{\centering
    \begin{rqbox}
        \textbf{RQ$_2$}: Is SelectQA more \textbf{efficient} than traditional state-of-the-art methods and than the BootQA method?
    \end{rqbox}
}

Since the metrics used to assess the efficiency and effectiveness of the classical and quantum strategies differ, the analysis is split into two parts.
First, we describe the research method followed to compare SelectQA to its classical counterparts and the results of this empirical analysis.
Then, we discuss how we compared SelectQA to BootQA and the results of this second analysis.

\subsection{Comparing SelectQA to Traditional TCS Algorithms}
\label{sec:c_comp}

%The motivations that justify the need for a quantum alternative to the classical algorithms for the resolution of the test case selection problems are different.
%Regression testing of large-scale real-world systems has to deal with vast and complex test suites, which also means intricate relationships between test cases; quantum algorithms have proven to be excellent at finding optimal solutions efficiently, even in complex spaces, potentially uncovering unknown correlations to the classical algorithms. Furthermore, quantum algorithms can simultaneously explore multiple configurations of the same complex space and incorporate additional constraints like memory limitation and specific coverage goals more effectively.

\begin{table}[ht]
\footnotesize
\caption{Characteristics of the programs under study}\label{table:sir_programs}
\begin{tabular}{@{}lrrr@{}} % Adjust the width of p{} as needed
\toprule
\textbf{Program} & \textbf{LOC} & \textbf{\# TCs} & \textbf{Description} \\
\midrule
flex    & 10,459   & 567  & Fast lexical analyzer  \\
grep    & 10,068   & 808  & Regular expression utility  \\
gzip    & 5,680    & 215  & Data compression program  \\
sed     & 14,427   & 360  & Non-interactive text editor  \\
\bottomrule
\end{tabular}
\end{table}

\subsubsection{Study Context}
One of the \textit{goals} of this study is to evaluate the performance of SelectQA in resolving the TCS problem in terms of the three objectives depicted earlier: code coverage (in particular \textit{statement coverage}), execution cost, and past faults coverage.
The \textit{study context} consists of four GNU open-source programs from the \textit{software-artifact infrastructure repository} (SIR)\cite{sir_rep}: \textit{flex}, \textit{grep}, \textit{gzip} and \textit{sed}.
\Cref{table:sir_programs} reports the main characteristics of such programs. The choice of these programs is not random since they have been used in previous work, including the one proposing DIV-GA~\cite{panichella2014improving}.
The following describes the test case selection criteria and how they have been gathered.

\begin{itemize}
\item \textit{Statement Coverage.} Statement coverage represents how test cases execute source code statements. To extract these data, we used \textit{gcov}, which can track the statements executed by each test case in C programs.

\item \textit{Execution Cost.} We did not rely on their execution time as external factors could influence it. Instead, the cost is calculated by counting the executed elementary instructions. This approach is consistent with previous work by Panichella et al.~\cite{panichella2014improving}. We used \textit{gcov} to determine the execution frequency of each basic block (i.e., a linear section of code that is not branched and has only one entry and exit point). 
Block count is preferred over line count, as one line may contain multiple branches or function calls.

\item \textit{Past Faults Coverage.} The SIR provides versions of the programs, each featuring \textit{injected} faults, specifying, through the use of a \textit{fault matrix}, whether a given test case has detected errors in the past. 
This information is translated into a binary value associated with the corresponding test cases.
\end{itemize}

%\subsubsection{\textbf{Research Questions}}
%\textit{In the context of the first branch of the comparisons}, this study aims to collect empirical evidence to answer:

%\begin{itemize}[label=$\cdot$]
%    \item \textbf{RQ\(_1\)} \textit{Is SelectQA more \textbf{effective} than the additional greedy, DIV-GA, and BootQA strategies?} This research question evaluates the ability of SelectQA to produce a large number of near-optimal solutions when compared to classical DIV-GA and additional greedy techniques.
%    \item \textbf{RQ\(_2\)} \textit{Is SelectQA more \textbf{efficient} than the additional greedy, DIV-GA, and BootQA strategies?} This research question aims to evaluate the performance of SelectQA in terms of execution times compared to DIV-GA and additional greedy techniques.
%\end{itemize}

\subsubsection{Experiment Configurations}
The compared algorithms have been executed ten times to ensure correct empirical analysis of the results and mitigate the effects of the random nature of quantum algorithms, such as the annealing used by SelectQA. We compared:

\begin{itemize}
    \item \textit{SelectQA} proposed in this paper, which resolves a QUBO reformulation of the TCS problem leveraging quantum annealing;
    \item \textit{DIV-GA} by Panichella et al. leverages diversity to generate more effective solutions~\cite{panichella2014improving};
    \item \textit{Additional Greedy} by Yoo and Harman~\cite{yoo2007pareto} unifies the three objectives into a single objective function to minimize and incrementally build a set of non-dominated solutions.
\end{itemize}

SelectQA and additional greedy have been implemented in Python, the former leveraging the \textit{dwave} and \textit{dimod} libraries.
To compare the three algorithms, we implemented the QUBO formalization for the three-objectives TCS problem (the results and the code to replicate the experiments follow the instructions in~\cite{appendix}). Since quantum annealing is a single-objective algorithm, we built its solution incrementally following the same strategy used by additional greedy. So, starting from the selected test cases obtained by the annealing process, we incrementally produced a set of non-dominated solutions (sub-suites)~\cite{appendix}.
DIV-GA has been implemented using \textit{MATLAB R2024a Global Optimization Toolbox}, customizing the \textit{gamultiobj} routine; the generation of the initial population has been performed using the \textit{rowexch}, \textit{hadamard} and \textit{sortrows} routines. We set the DIV-GA parameters as described in the original work~\cite{panichella2014improving}.
All implementations are available as part of our online appendix~\cite{appendix}.

\subsubsection{Evaluation Metrics}
\label{sec:4.7}
\noindent\textit{Effectiveness.} To evaluate the quality of a multi-objective optimization algorithm, its yielded Pareto frontier should be compared to the actual one. Nevertheless, when resolving large problems like TCS, knowing the actual Pareto frontier \textit{a priori} is impossible due to the impossibility of an exhaustive search.
The only way is to perform an \textit{a posteriori} evaluation, building, following the work by Panichella et al.~\cite{panichella2014improving}, a hybrid frontier composed of all the non-dominated solutions between all the different frontiers obtained by each different algorithm for all the runs. Such a hybrid frontier is called \textit{reference Pareto frontier}~\cite{yoo2007pareto}.
Let $P=\{P_1,...,P_l\}$ be the set of $l$ different Pareto frontiers obtained after all the experiment runs by all the evaluated algorithms, the Pareto frontier $P_{ref}$ is defined as follows.

\begin{align}
P_{ref} \subseteq \bigcup_{i=1}^{l} P_i
\end{align}

where $\forall p \in P_{ref} \nexists q \in P_{ref} : q > p$.

Given the reference frontier, we computed the \textit{number of non-dominated solutions}, i.e., the number of non-dominated solutions found by an algorithm selected for the final reference frontier. 

To ensure the empirical reliability of the results, we statistically analyzed the results to check whether the differences between the results obtained by the compared algorithms are significant. The results obtained over ten independent runs have been compared using \textit{Mann-Whitney U test}~\cite{t_test}. Significant $p$-values mean that the null hypothesis, i.e., there is no statistically relevant difference between the effectiveness of the algorithms, has to be rejected in favor of the alternative one: statistically speaking, one of the two algorithms exhibits more non-dominated solutions selected by the reference Pareto frontier (we reject the null hypothesis for $p$-values $<$ 0.05).
Finally, we estimated the magnitude of the difference between the performance reported by the algorithms using the \textit{Vargha-Delaney effect size} (\^A$_{12}$)~\cite{a12}. The effect size is interpreted as \textit{small} in the range $(0.34,0.44]$ or $[0.56,0.64)$, \textit{medium} in the range $(0.29, 0.34]$ or $[0.64,0.71)$, and \textit{large} in the range $[0,0.29]$ or $[0.71,1]$.

\noindent\textit{Efficiency.} We analyzed the mean total run time of the algorithms over ten independent runs to compare their efficiency.
DIV-GA and additional greedy have been executed on an Apple Macbook Air featuring an M1 chip and 16GB of RAM.
The quantum annealing process was run on the D-Wave Leap Hybrid Solver \texttt{hybrid\_binary\_quadratic\_model\_version2}~\cite{tech_rep}. We used the D-Wave's ``run-time'' metric to obtain reliable results, reporting the total time needed for the machine to finish the operation.
We statistically analyzed the total run times obtained over ten independent runs by each of the three algorithms leveraging the \textit{Mann-Whitney U test}. We quantified the magnitudes of their differences using the \textit{Vargha-Delaney effect size} (\^A$_{12}$).

\subsubsection{Results: RQ1 - Effectiveness}

\begin{table*}[ht]
    \centering
    \footnotesize
    \caption{Mean Pareto size and number of non-dominated solutions obtained on average by the algorithms in ten runs}
    \begin{tabular}{llrrrr}
        \toprule
        \multirow{2}{*}{Program} & \multirow{2}{*}{Method} & \multicolumn{2}{c}{Pareto Size} & \multicolumn{2}{c}{Non-Dom Solutions} \\
        \cmidrule(r){3-4} \cmidrule(r){5-6}
        &  & Mean & St. Dev. & Mean & St. Dev. \\
        \midrule
        flex & SelectQA & 187.0 & - & 187.0 & - \\
        & DIV-GA & 140.0 & - & 140.0 & - \\
        & Add. Greedy & \textbf{567.0} & - & \textbf{205.0} & - \\
        & Additional Method & 150.0 & 2.5 & 150.0 & 2.5 \\
        \midrule
        grep & SelectQA & 225.5 & 0.5 & \textbf{207.5} & 0.52 \\
        & DIV-GA & 70.0 & - & 70.0 & - \\
        & Add. Greedy & \textbf{802.0} & - & 177.0 & - \\
        & Additional Method & 180.0 & 3.0 & 180.0 & 3.0 \\
        \midrule
        gzip & SelectQA & 86.3 & 0.8 & 41.3 & 0.8 \\
        & DIV-GA & 105.0 & - & \textbf{105.0} & - \\
        & Add. Greedy & \textbf{199.0} & - & 71.0 & - \\
        & Additional Method & 95.0 & 1.5 & 95.0 & 1.5 \\
        \midrule
        sed & SelectQA & 131.0 & - & \textbf{131.0} & - \\
        & DIV-GA & 99.6 & 13.4 & 99.6 & 13.4 \\
        & Add. Greedy & \textbf{356.0} & - & 85.0 & - \\
        & Additional Method & 120.0 & 5.0 & 120.0 & 5.0 \\
        \bottomrule
    \end{tabular}
    \label{table:rq1_results}
\end{table*}

\begin{table*}[ht]
    \centering
    \footnotesize
    \caption{Statistical comparisons between the algorithms in terms of the number of non-dominated solutions}
    \begin{tabular}{llrr}
        \toprule
        \multirow{2}{*}{\textbf{Program}} & \multirow{2}{*}{\textbf{Hypothesis}} & \multicolumn{2}{c}{\textbf{Non-Dom Solutions}} \\
        \cmidrule(r){3-4}
         &  & \textbf{p-value} & \textbf{\^A$_{12}$} \\
        \midrule
        flex & SelectQA$>$DIV-GA & \textbf{$<$0.01} & 1.0 (L) \\
        & Add. Greedy$>$SelectQA & \textbf{$<$0.01} & 1.0 (L) \\
        & Add. Greedy$>$DIV-GA & \textbf{$<$0.01} & 1.0 (L) \\
        \midrule
        grep & SelectQA$>$DIV-GA & \textbf{$<$0.01} & 1.0 (L) \\
        & SelectQA$>$Add. Greedy & \textbf{$<$0.01} & 1.0 (L) \\
        & Add. Greedy$>$DIV-GA & \textbf{$<$0.01} & 1.0 (L) \\
        \midrule
        gzip & DIV-GA$>$SelectQA & \textbf{$<$0.01} & 1.0 (L) \\
        & DIV-GA$>$Add. Greedy & \textbf{$<$0.01} & 1.0 (L) \\
        & Add. Greedy$>$SelectQA & \textbf{$<$0.01} & 1.0 (L) \\
        \midrule
        sed & SelectQA$>$DIV-GA & \textbf{$<$0.01} & 1.0 (L) \\
        & SelectQA$>$Add. Greedy & \textbf{$<$0.01} & 1.0 (L) \\
        & DIV-GA$>$Add. Greedy & \textbf{$<$0.01} & 0.9 (L) \\
        \bottomrule
    \end{tabular}
    \label{table:rq1_stats}
\end{table*}

\Cref{table:rq1_results} reports the means and standard deviations of the size of the Pareto frontier and the number of non-dominated solutions for the three compared algorithms obtained by executing them ten times independently.
The additional greedy algorithm always finds the larger set of solutions, but those of the other algorithms generally dominate those solutions. Still, additional greedy performed better than the other approaches for the flex program.
SelectQA is the most effective approach, finding the highest number of non-dominated solutions among all the compared algorithms on two out of four programs. Concerning gzip, DIV-GA performs better than SelectQA and additional greedy. DIV-GA appears to be the most precise approach because all the solutions it finds are selected by the reference frontier.

To support the results, \Cref{table:rq1_stats} reports the results of the Mann-Whitney U test and the Vargha-Delaney \^A$_{12}$ effect size, obtained comparing the number of non-dominated solutions obtained by the three algorithms in the ten different runs. The test results confirm our previous observations. SelectQA performs statistically better than all the other algorithms in two of four programs, always with a large magnitude. In the case of gzip, DIV-GA performs statistically better than the other algorithms with large magnitude; the same is true for additional greedy regarding flex. Also, additional greedy performed better than SelectQA for gzip.

\begin{figure*}[!ht]
    \centering
    % Subfigure 1
    \subfigure[Flex]{\includegraphics[width=0.4\textwidth]{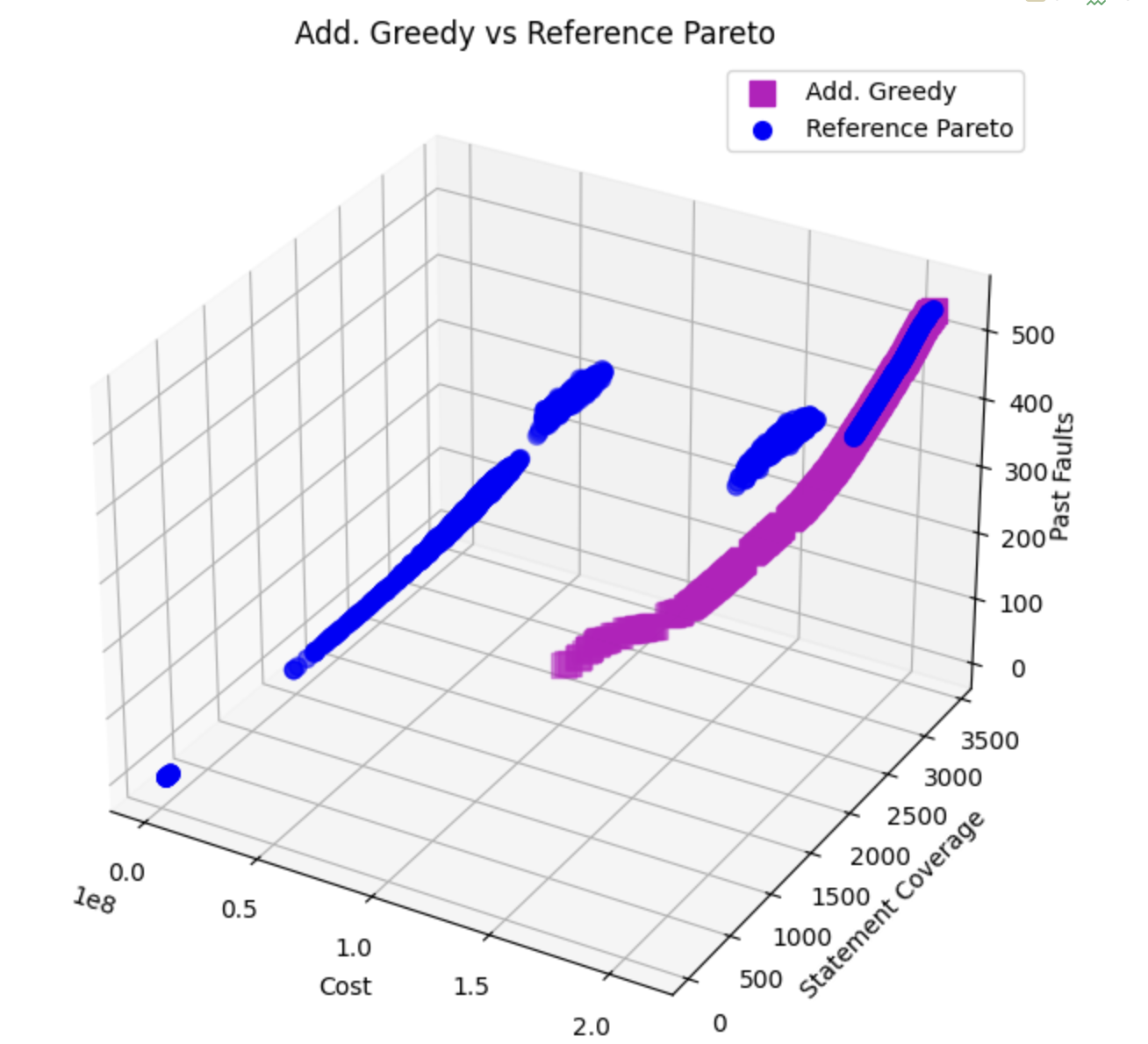}}
    % Subfigure 2
    \subfigure[Grep]{\includegraphics[width=0.4\textwidth]{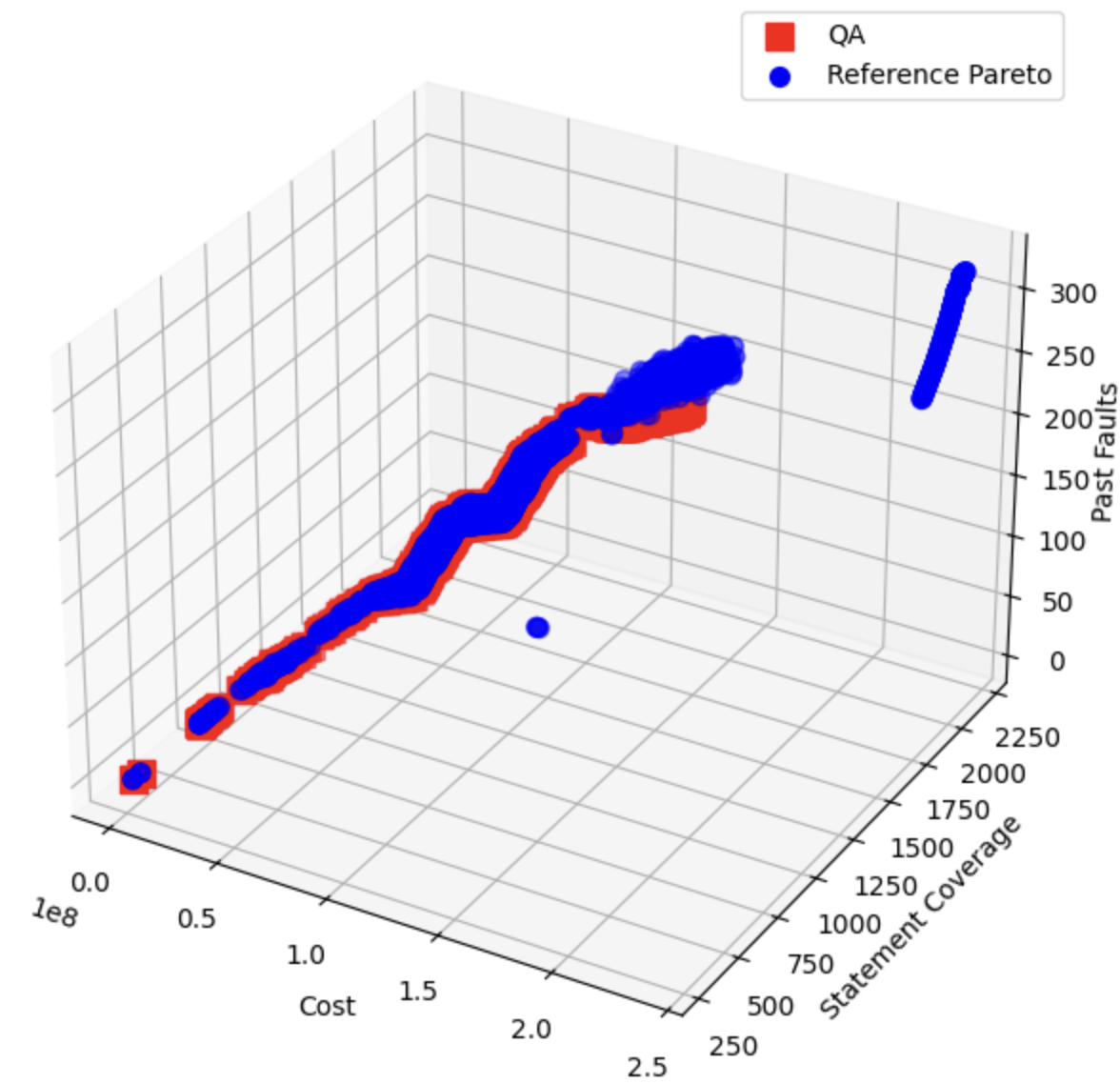}}
    % Subfigure 3
    \subfigure[Gzip]{\includegraphics[width=0.4\textwidth]{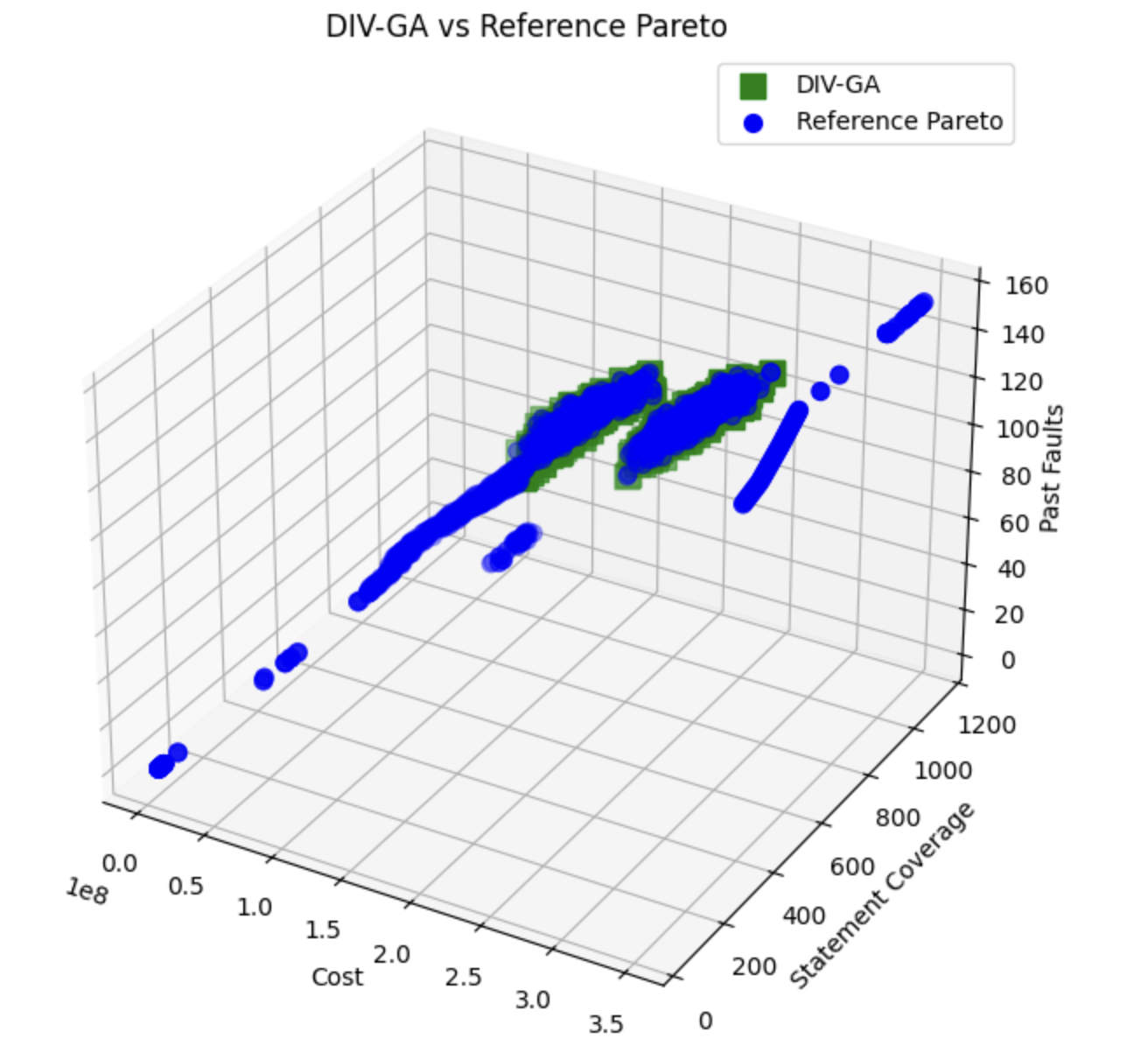}}
    % Subfigure 4
    \subfigure[Sed]{\includegraphics[width=0.4\textwidth]{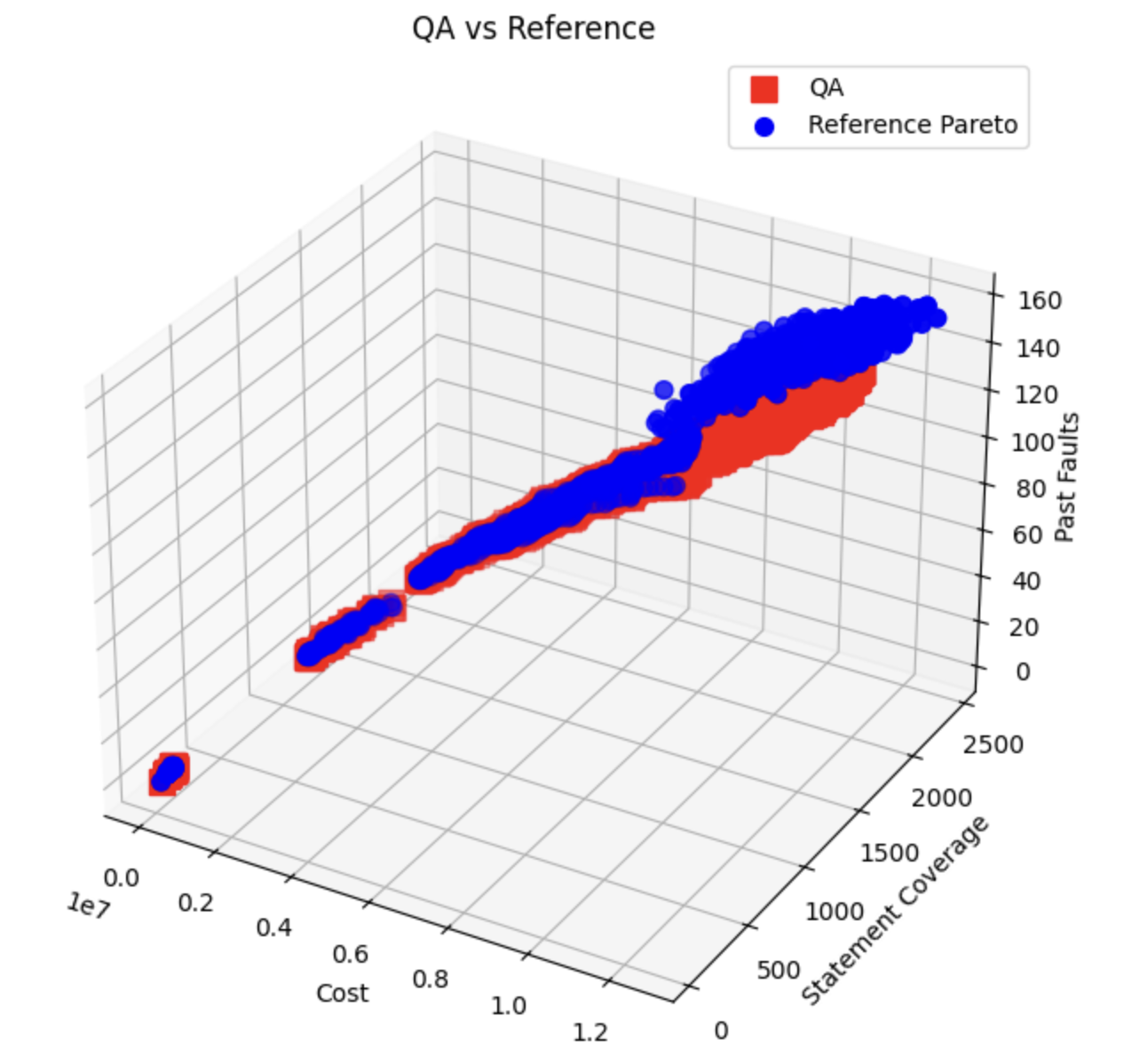}}
    \caption{Additional Greedy performed the best with flex, while SelectQA with grep and sed, and DIV-GA with gzip.} % Figure caption
    \label{fig:pareto_frontiers} % Label for referencing
\end{figure*}

\begin{figure}[!ht]
    \centering
    
\end{figure}

\Cref{fig:pareto_frontiers} provides, for each program, a graphical comparison between the frontier obtained by a single execution of the best algorithm for that program and the corresponding reference Pareto frontier.
The figure is consistent with the results reported in the previous section. Except for gzip and flex, SelectQA can always find the largest number of non-dominated solutions; we can see its frontier covering the larger part of the reference frontiers of grep and sed. DIV-GA is very precise: its solutions are always on the reference frontier, although generally, they are less than those obtained by SelectQA.

{\centering
    \begin{rqbox}
        \textbf{Takeaway \#1.}
        SelectQA is the most effective approach in the number of non-dominated solutions, whereas DIV-GA remains unbeaten in the percentage of non-dominated solutions found.
    \end{rqbox}
}

\subsubsection{Results RQ2 - Efficiency}

\begin{table}[!ht]
    \centering
    \footnotesize
    \caption{Average execution times and standard deviations of the algorithms}
    \label{table:rq2_results}
    \begin{tabular}{lrrrrrr}
        \toprule
        & \multicolumn{2}{c}{\textbf{Add. Greedy}} & \multicolumn{2}{c}{\textbf{DIV-GA}} & \multicolumn{2}{c}{\textbf{SelectQA}}\\
        \cmidrule(lr){2-3} \cmidrule(lr){4-5}  \cmidrule(lr){6-7}
        & \textbf{Mean} & \textbf{SD} & \textbf{Mean} & \textbf{SD} & \textbf{Mean} & \textbf{SD} \\
        \midrule
        flex & 8s & - & 3m 39s & 14s & \textbf{2.9s} & 0.003s \\
        grep & 8s & - & 1m 27s & 5s & \textbf{2.9s} & 0.003s \\
        gzip & \textbf{$<$ 1s} & - & 19s & 2s & 2.9s & 0.003s \\
        sed & \textbf{1s} & - & 1m 20s & 8s & 2.9s & 0.003s \\
        \bottomrule
    \end{tabular}
\end{table}

\Cref{table:rq2_results} shows that SelectQA can resolve the problems constantly. Due to the use of singular value decomposition~\cite{panichella2014improving}, DIV-GA is quite expensive and requires more computational time (compensated by the quality of its solutions). So, SelectQA is always more efficient than DIV-GA.

Additional greedy performed better than SelectQA in two cases: the gzip and sed programs. In all the other cases, SelectQA performs better than additional greedy. Please note that the additional greedy algorithm has a computational time of $O(|T| \cdot max|T_i|)$, in which $|T|$ is the size of the starting test suite and $max|T_i|$ represents the cardinality of the largest set of test cases able to reach the maximum coverage. The larger the system, the larger $max|T_i|$ will be, with high values leading to a higher number of iterations for the algorithm (because the maximum coverage to reach could be very high), meaning that applying additional greedy would be impractical for larger systems.

The statistical analysis confirmed the findings reported in \Cref{table:rq2_results}, showing p-values always smaller than 0.01 with large effect sizes.

{\centering
    \begin{rqbox}
        \textbf{Takeaway \#2.}
        SelectQA has constant execution time despite the size of the program under test and outperforms DIV-GA in efficiency. Additional greedy could be impractical for large systems.
    \end{rqbox}
}

\subsection{Comparing SelectQA to BootQA}
\label{sec5:q_comp}
\subsubsection{Study Context}
The \textit{context} of this work consists of two real-world datasets, used in the previous work introducing BootQA~\cite{bootqa}:\textit{PaintControl} from ABB Robotics Norway~\cite{paintcontrol_ds} and \textit{GSDTSR}~\cite{gsdtsr} from Google. Both these datasets have the properties: ``execution time'' and ``failure rate'' (coherently with~\cite{bootqa}, we filtered out the test cases of both datasets with failure rates at zero, i.e., tests that never triggered failures during their history).
We adapted SelectQA by applying the two-objective QUBO formulation presented in \Cref{sec:3.2.2}, where the two criteria are execution time and failure rate.

%\subsubsection{\textbf{Research Questions}}
%\textit{In the context of the second branch of the comparisons}, this study aims to collect empirical evidence to answer:

%\begin{itemize}[label=$\cdot$]
%    \item \textbf{RQ\(_1\)} \textit{Is SelectQA more \textbf{effective} than the additional greedy, DIV-GA, and BootQA strategies?} 
%    This research question evaluates the ability of SelectQA to produce a large number of near-optimal solutions when compared to BootQA.
%    \item \textbf{RQ\(_2\)} \textit{Is SelectQA more \textbf{efficient} than the additional greedy, DIV-GA, and BootQA strategies?}
%    This research question aims to evaluate the performance of SelectQA in terms of \textit{qpu execution time} compared to BootQA.
%\end{itemize}

\subsubsection{\textbf{Experiment Configuration}}
The compared algorithms have been executed ten independent times to ensure correct empirical analysis of the results, especially since both tools rely on quantum algorithms.

\begin{table}[ht]
    \centering
    \caption{Best BootQA configuration for each dataset}

    \begin{tabular}{lrrr}
        \toprule
        \textbf{Dataset} & \textbf{\# TCs} & \textbf{n} & \textbf{m}\\
        \midrule
        PaintControl & 89 & 30 & 6 \\
        GSDTSR & 287 & 20 & 21 \\
        \bottomrule
    \end{tabular}
    \label{table:bootqa_datasets}

\end{table}

BootQA was directly cloned from its public repository in GitHub\cite{bootqa_rep}.
For both PaintControl and GSDTSR datasets, ten independent executions were conducted to empirically evaluate the best configurations of ($n,m$) parameters. Afterward, the algorithm was run ten independent times on the two datasets with the optimal configurations. \Cref{table:bootqa_datasets} reports the best configurations found for GSDTSR and PaintControl, coherent with the base configuration~\cite{bootqa}.
%The version of SelectQA, showed in Listing \ref{lst:qubo_generation2}, implements the two-objective formalization of the QUBO representation of the test case selection problem presented in Section \ref{sec:3.2.2}. %This choice has been made to run the BootQA tool on the same dataset and with the same conditions described in its publication paper\cite{bootqa}.
As previously described, SelectQA has been implemented in Python, using the \textit{dwave} and \textit{dimod} libraries.
All implementations are available as part of our online appendix~\cite{appendix}.

\subsubsection{\textbf{Evaluation Metrics}}
\label{sec:5.7}

\noindent\textit{Effectiveness.} Since both SelectQA and BootQA use a single-objective formulation to obtain a sub-optimal test suite, we compared the solutions by evaluating the execution times and failure rates of the test suites obtained by the two strategies.
Coherently with the previous work on BootQA~\cite{bootqa}, we statistically analyzed the results obtained over ten independent executions by applying the \textit{Mann-Whitney U test}~\cite{t_test} with a \textit{significance level} set at 0.05. The null hypothesis represents a non-relevant difference between the two approaches. In contrast, if the null hypothesis is rejected, the magnitude of the difference is computed using the \textit{Vargha-Delaney} effect size (\^A$_{12}$)~\cite{a12} (we reject the null hypothesis for $p$-values $<$ 0.05).
The effect size is interpreted as \textit{small} in the range $(0.34,0.44]$ or $[0.56,0.64)$, \textit{medium} in the range $(0.29, 0.34]$ or $[0.64,0.71)$, and \textit{large} in the range $[0,0.29]$ or $[0.71,1]$.

\noindent\textit{Efficiency.} 
To compare the efficiency of BootQA and SelectQA, we analyzed their total run times.
In particular, the former executes a local decomposition through bootstrap sampling (in this work, executed on a MacBook Air featuring an M1 Chip and 16GB of RAM) and directly runs the \textit{Advanced System QPU}. The latter only relies on the {\texttt{hybrid\_binary\_quadratic\_model\_version2}} to handle highly complex optimization problems. For BootQA, we considered the total run times to be the sum of the bootstrap sampling process and Advanced System QPU execution time. In contrast, for SelectQA, we computed the total run time of the {\texttt{hybrid\_binary\_quadratic\_model\_version2}}. The empirical reliability of the findings has been statistically validated by analyzing the distribution of the total run times, obtained over ten independent runs by each algorithm, using the \textit{Mann-Whitney U test}. The magnitudes of the differences between the sequences have been quantified using the \textit{Vargha-Delaney effect size} (\^A$_{12}$).

\subsubsection{Results: RQ1 - Effectiveness}
\label{sec:req1}

\begin{figure}[ht] % Place the figure at the top of the page
    \centering
    
    % Subfigure 1
    \subfigure[PaintControl]{\includegraphics[width=0.35\textwidth]{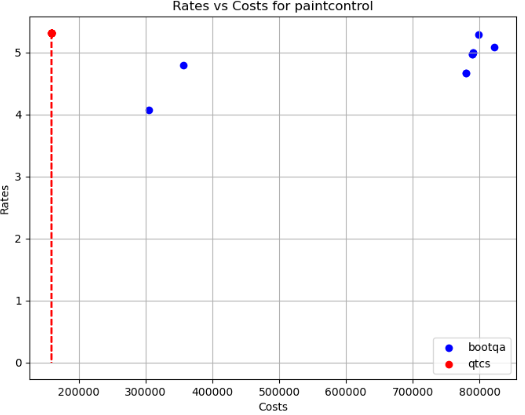}}\hfill
    % Subfigure 2
    \subfigure[GSDTSR]{\includegraphics[width=0.35\textwidth]{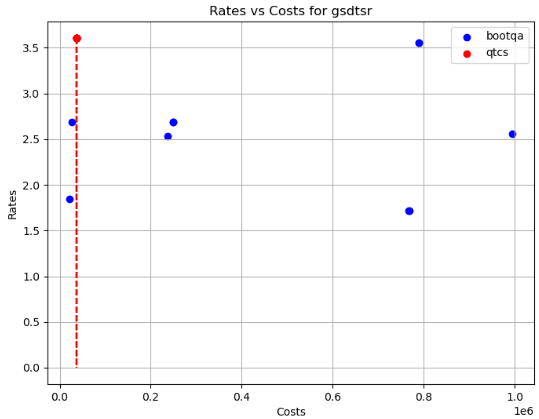}} \\
    
    \caption{Costs and failure rates comparisons} % Figure caption
    \label{fig:datasets_comparisons} % Label for referencing
\end{figure}

As seen in \Cref{fig:datasets_comparisons}, SelectQA could find the optimal solution to the problem in each run and for both datasets. The $\alpha$ parameter, balancing the weight of failure rate and execution time, can direct SelectQA to the solution representing the optimal trade-off between the two research objectives. It also allows engineers to decide whether one objective should be more relevant than another.
BootQA could not find a solution that dominates SelectQA in terms of execution times and failure rates, while SelectQA found a solution that dominates those found by BootQA on 18 runs out of 20.
In the remaining two cases on the GSDTSR dataset, BootQA could find two test suites with better execution times than those found by SelectQA but with worse failure rates; hence, BootQA could never dominate the solutions found by SelectQA.

\begin{table}[ht]
    \centering
    \caption{Statistical Comparisons between the Algorithms}
    \footnotesize
    \begin{tabular}{cccc}
        \toprule
        \multirow{2}{*}{\textbf{Dataset}} & \multirow{2}{*}{\textbf{Hypothesis}} & \multirow{2}{*}{\textbf{p-value}} & \multirow{2}{*}{\textbf{\^A$_{12}$}}\\
        \\
        \midrule
        PaintControl & SelectQA$<$BootQA costs & \textbf{$<$0.01} & 0.0(L) \\
         & SelectQA$>$BootQA f rate & \textbf{$<$0.01} & 1.0(L) \\
        GSDTSR & SelectQA$<$BootQA costs & \textbf{0.02} & 0.2(L) \\
         & SelectQA$>$BootQA f rate & \textbf{$<$0.01} & 1.0(L) \\
        \bottomrule
    \end{tabular}
    \label{table:rq3_stats}
\end{table}

To understand the magnitude of the difference between the two approaches, we report in \Cref{table:rq3_stats} the results of the Mann-Whitney U test and the Vargha and Delaney's \^A$_{12}$ effect size, obtained by comparing the values of execution times and failure rates of the test suites obtained by the two approaches (over all the ten runs). The test results confirm the previous observations and point out a large magnitude of the difference between the two approaches in favor of SelectQA.

{\centering
    \begin{rqbox}
        \textbf{Takeaway \#3.}
        SelectQA outperforms BootQA in effectiveness, always finding the optimal trade-off solution and dominating BootQA solutions in most cases.
    \end{rqbox}
}

\subsubsection{Results: RQ2 - Efficiency}
\label{sec:rq2}

\begin{table}[ht]
    \centering
    \caption{Average Execution Times of the Compared Algorithms}
    \footnotesize
    \label{table:rq4_results}
    \begin{tabular}{lrrrr}
        \toprule
        & \multicolumn{2}{c}{\textbf{SelectQA}} & \multicolumn{2}{c}{\textbf{BootQA}} \\
        \cmidrule(lr){2-3} \cmidrule(lr){4-5}
        & \textbf{Mean} & \textbf{SD} & \textbf{Mean} & \textbf{SD} \\
        \midrule
        PaintControl & 2.9s & 0.003s & \textbf{0.029s} & $<$0.001s \\
        GSDTSR & 2.9s & 0.003s & \textbf{0.030s} & $<$0.001s \\
        \bottomrule
    \end{tabular}
\end{table}

As expected, \Cref{table:rq4_results} shows that BootQA is more efficient than SelectQA regarding the mean total run time because the problems to solve in BootQA are way smaller than in SelectQA. Also, the statistical analysis confirmed the finding, showing all the p-values as smaller than 0.01 with large 0 effect sizes.

{\centering
    \begin{rqbox}
        \textbf{Takeaway \#4.}
        BootQA outperforms SelectQA in total run time, showing the efficiency of a pure quantum solution.
    \end{rqbox}
}

\section{Threats to Validity}
\label{sec6:ttv}

This section discusses all aspects that could threaten the validity of the study conducted about the comparisons between SelectQA, DIV-GA, additional greedy, and BootQA.

\begin{itemize}
\item \textit{Construct Validity.} The main threat in this regard concerns the correctness of the measures used to select tests: coverage, history of failures, cost of execution, and failure rate. We relied on open-source profiling and compilation tools (e.g., GNU gcc and gcov) and real-world case studies to limit this issue.
DIV-GA has been implemented in MATLAB, while additional greedy and SelectQA have been implemented in Python. This difference could threaten validity due to the different code optimization mechanisms used by the two languages and the routines applied by the  MATLAB environment to solve mathematical optimization problems. However, we performed this choice to replicate the DIV-GA base conditions and environment as presented in~\cite{panichella2014improving}.
The choice of efficiency metric constitutes another threat to construct validity. We have chosen the total run time to compare SelectQA with the traditional and quantum algorithms. Such a metric provides a straightforward measure of how long the examined strategies take to solve the requested problem; however, this choice poses some significant challenges. The SelectQA total run time not only considers the time used by CPUs, like for its classic counterparts, but also the execution time of the Advantage QPU; hence, the total run time reported by SelectQA could be subject to considerable variability depending on QPU availability, resource queues, and classical-quantum communication overhead. In other words, the total run time of SelectQA could be subject to different bottlenecks due to various causes.

\item \textit{Internal Validity.}
One of these threats is undoubtedly the random nature of quantum annealing algorithms. 
For this reason, the experiments were repeated ten times for each program under examination and then considering the means of the obtained results.
The \textit{tuning} of the $P$ penalty parameter is also a factor that could undermine the internal validity of this job. Therefore, we applied a method well-known in literature~\cite{penalty_weights} when choosing this value.
The $\alpha$ parameter has been validated following repeated trials.
We used the D-Wave default settings for other QA hyper-parameters configurations coherently with the original BootQA paper~\cite{bootqa} and also asked the researchers themselves who worked on it about the parameters' tuning.
Another threat to internal validity concerns the implementation of DIV-GA. The original version used the MATLAB \textit{Global Optimization Toolbox} release R2011b, while the version used for this work is the MATLAB \textit{Global Optimization Toolbox} release R2024a. So, the new version of DIV-GA could be subject to differences compared to its older version due to updated routines and different code optimization mechanisms.
Also, the difference between the languages used to implement DIV-GA, additional greedy, and SelectQA could be seen as a threat to the internal validity due to the aforementioned fluctuations. This threat has been mitigated by executing different runs for each algorithm using the same datasets.
Finally, another threat to internal validity is that using other sampling strategies for the ${(m,n)}$ parameters of BootQA might result in better solutions. Our choice was to replicate the experiment conditions of the original BootQA work~\cite{bootqa}, so we did not experiment with different sampling strategies.

\item \textit{External Validity.}
A dangerous threat to external validity corresponds to the impossibility of generalizing the obtained results, which is particularly true if the datasets are too small or distant from real-world scenarios. To mitigate this issue, the datasets chosen for the comparisons, extracted from reliable sources such as SIR, Google, and ABB Robotics Norway, have already been used for evaluating the approaches used as baselines in our study~\cite{yoo2007pareto}~\cite{yoo2010using}~\cite{yoo2011highly}~\cite{bootqa}~\cite{panichella2014improving}. The chosen SIR programs represent real-world medium-scale and small-scale scenarios. On the other hand, the datasets used to compare SelectQA and BootQA represent real-world industrial medium-scale and small-scale datasets.

\item \textit{Conclusion Validity.}
We interpreted the findings using appropriate statistical tests. To test the significance of the differences, we applied the (i) \textit{Mann-Whitney U test}~\cite{t_test}, while to estimate the magnitude and the effect size of the observed differences, we used the \textit{Vargha-Delaney}~\cite{t_test} (\^A$_{12}$) effect size. Conclusions are based only on statistically significant results.
\end{itemize}

\section{Conclusions}
\label{sec:conclusion}
This paper proposes SelectQA, a method for test case selection leveraging quantum annealing, and compares it to classical and quantum approaches.
Regarding the classical approaches~\cite{yoo2007pareto,panichella2014improving}, we found that it is the most effective approach in the number of non-dominated solutions, whereas DIV-GA~\cite{panichella2014improving}, able to beat SelectQA in just one case, remains unbeaten in the percentage of non-dominated solutions found.
Looking at the efficiency of the algorithms, SelectQA performs in constant time, demonstrating the superiority of quantum algorithms over classical ones due to the program size-independent execution time.
Regarding BootQA~\cite{bootqa}, i.e., another method based on quantum computing, we found that the solutions provided by SelectQA dominate those offered by the other in most cases, whereas the opposite never happens.
Nevertheless, BootQA is the most efficient algorithm due to the smaller size of problems deployed directly to the quantum annealer.

In future work, we plan to implement additional problem decomposition strategies with quantum algorithms to obtain more precise, stable, and efficient solutions. We also aim to consider other quantum algorithms in the experiments; in particular, we will implement a novel QAOA strategy to further compare classical, annealing, and QAOA strategies.

\section*{Acknowledgement}
This work has been partially supported by the project ``RECHARGE: monitoRing, dEtection, and CHaracterization of performAnce ReGrEssionss'', grant P2022SELA7, under the PRIN 2022 PNRR MUR program funded by the EU - NGEU.

This work has been partially supported by the project ``QUASAR: QUAntum software engineering for Secure, Affordable, and Reliable systems'', grant 2022T2E39C, under the PRIN 2022 MUR program funded by the EU - NGEU.

Views and opinions expressed are, however, those of the author(s) only and do not necessarily reflect those of the European Union or The European Research Executive Agency. Neither the European Union nor the granting authority can be held responsible.

This version of the article has been accepted for publication, after peer review (when applicable) and is subject to Springer Nature’s AM terms of use, but is not the Version of Record and does not reflect post-acceptance improvements, or any corrections. The Version of Record is available online at: http://dx.doi.org/10.1007/s10009-024-00775-w

%%===========================================================================================%%
%% If you are submitting to one of the Nature Portfolio journals, using the eJP submission   %%
%% system, please include the references within the manuscript file itself. You may do this  %%
%% by copying the reference list from your .bbl file, paste it into the main manuscript .tex %%
%% file, and delete the associated \verb+\bibliography+ commands.                            %%
%%===========================================================================================%%

\bibliography{sn-bibliography}% common bib file

%% BioMed_Central_Bib_Style_v1.01

\begin{thebibliography}{45}
% BibTex style file: bmc-mathphys.bst (version 2.1), 2014-07-24
\ifx \bisbn   \undefined \def \bisbn  #1{ISBN #1}\fi
\ifx \binits  \undefined \def \binits#1{#1}\fi
\ifx \bauthor  \undefined \def \bauthor#1{#1}\fi
\ifx \batitle  \undefined \def \batitle#1{#1}\fi
\ifx \bjtitle  \undefined \def \bjtitle#1{#1}\fi
\ifx \bvolume  \undefined \def \bvolume#1{\textbf{#1}}\fi
\ifx \byear  \undefined \def \byear#1{#1}\fi
\ifx \bissue  \undefined \def \bissue#1{#1}\fi
\ifx \bfpage  \undefined \def \bfpage#1{#1}\fi
\ifx \blpage  \undefined \def \blpage #1{#1}\fi
\ifx \burl  \undefined \def \burl#1{\textsf{#1}}\fi
\ifx \doiurl  \undefined \def \doiurl#1{\url{https://doi.org/#1}}\fi
\ifx \betal  \undefined \def \betal{\textit{et al.}}\fi
\ifx \binstitute  \undefined \def \binstitute#1{#1}\fi
\ifx \binstitutionaled  \undefined \def \binstitutionaled#1{#1}\fi
\ifx \bctitle  \undefined \def \bctitle#1{#1}\fi
\ifx \beditor  \undefined \def \beditor#1{#1}\fi
\ifx \bpublisher  \undefined \def \bpublisher#1{#1}\fi
\ifx \bbtitle  \undefined \def \bbtitle#1{#1}\fi
\ifx \bedition  \undefined \def \bedition#1{#1}\fi
\ifx \bseriesno  \undefined \def \bseriesno#1{#1}\fi
\ifx \blocation  \undefined \def \blocation#1{#1}\fi
\ifx \bsertitle  \undefined \def \bsertitle#1{#1}\fi
\ifx \bsnm \undefined \def \bsnm#1{#1}\fi
\ifx \bsuffix \undefined \def \bsuffix#1{#1}\fi
\ifx \bparticle \undefined \def \bparticle#1{#1}\fi
\ifx \barticle \undefined \def \barticle#1{#1}\fi
\bibcommenthead
\ifx \bconfdate \undefined \def \bconfdate #1{#1}\fi
\ifx \botherref \undefined \def \botherref #1{#1}\fi
\ifx \url \undefined \def \url#1{\textsf{#1}}\fi
\ifx \bchapter \undefined \def \bchapter#1{#1}\fi
\ifx \bbook \undefined \def \bbook#1{#1}\fi
\ifx \bcomment \undefined \def \bcomment#1{#1}\fi
\ifx \oauthor \undefined \def \oauthor#1{#1}\fi
\ifx \citeauthoryear \undefined \def \citeauthoryear#1{#1}\fi
\ifx \endbibitem  \undefined \def \endbibitem {}\fi
\ifx \bconflocation  \undefined \def \bconflocation#1{#1}\fi
\ifx \arxivurl  \undefined \def \arxivurl#1{\textsf{#1}}\fi
\csname PreBibitemsHook\endcsname

%%% 1
\bibitem[\protect\citeauthoryear{Yoo and Harman}{2012}]{yoo2012regression}
\begin{barticle}
\bauthor{\bsnm{Yoo}, \binits{S.}},
\bauthor{\bsnm{Harman}, \binits{M.}}:
\batitle{Regression testing minimization, selection and prioritization: a
  survey}.
\bjtitle{Software testing, verification and reliability}
\bvolume{22}(\bissue{2}),
\bfpage{67}--\blpage{120}
(\byear{2012})
\end{barticle}
\endbibitem

%%% 2
\bibitem[\protect\citeauthoryear{Rothermel and
  Harrold}{1998}]{rothermel1998empirical}
\begin{barticle}
\bauthor{\bsnm{Rothermel}, \binits{G.}},
\bauthor{\bsnm{Harrold}, \binits{M.J.}}:
\batitle{Empirical studies of a safe regression test selection technique}.
\bjtitle{IEEE Transactions on Software Engineering}
\bvolume{24}(\bissue{6}),
\bfpage{401}--\blpage{419}
(\byear{1998})
\end{barticle}
\endbibitem

%%% 3
\bibitem[\protect\citeauthoryear{Perrouin et~al.}{2012}]{perrouin2012pairwise}
\begin{barticle}
\bauthor{\bsnm{Perrouin}, \binits{G.}},
\bauthor{\bsnm{Oster}, \binits{S.}},
\bauthor{\bsnm{Sen}, \binits{S.}},
\bauthor{\bsnm{Klein}, \binits{J.}},
\bauthor{\bsnm{Baudry}, \binits{B.}},
\bauthor{\bsnm{Le~Traon}, \binits{Y.}}:
\batitle{Pairwise testing for software product lines: comparison of two
  approaches}.
\bjtitle{Software Quality Journal}
\bvolume{20},
\bfpage{605}--\blpage{643}
(\byear{2012})
\end{barticle}
\endbibitem

%%% 4
\bibitem[\protect\citeauthoryear{Yoo and Harman}{2007}]{yoo2007pareto}
\begin{bchapter}
\bauthor{\bsnm{Yoo}, \binits{S.}},
\bauthor{\bsnm{Harman}, \binits{M.}}:
\bctitle{Pareto efficient multi-objective test case selection}.
In: \bbtitle{Proceedings of the 2007 International Symposium on Software
  Testing and Analysis},
pp. \bfpage{140}--\blpage{150}
(\byear{2007})
\end{bchapter}
\endbibitem

%%% 5
\bibitem[\protect\citeauthoryear{Yoo and Harman}{2010}]{yoo2010using}
\begin{barticle}
\bauthor{\bsnm{Yoo}, \binits{S.}},
\bauthor{\bsnm{Harman}, \binits{M.}}:
\batitle{Using hybrid algorithm for pareto efficient multi-objective test suite
  minimisation}.
\bjtitle{Journal of Systems and Software}
\bvolume{83}(\bissue{4}),
\bfpage{689}--\blpage{701}
(\byear{2010})
\end{barticle}
\endbibitem

%%% 6
\bibitem[\protect\citeauthoryear{Yoo et~al.}{2011}]{yoo2011highly}
\begin{bchapter}
\bauthor{\bsnm{Yoo}, \binits{S.}},
\bauthor{\bsnm{Harman}, \binits{M.}},
\bauthor{\bsnm{Ur}, \binits{S.}}:
\bctitle{Highly scalable multi objective test suite minimisation using graphics
  cards}.
In: \bbtitle{Search Based Software Engineering: Third International Symposium,
  SSBSE 2011, Szeged, Hungary, September 10-12, 2011. Proceedings 3},
pp. \bfpage{219}--\blpage{236}
(\byear{2011}).
\bcomment{Springer}
\end{bchapter}
\endbibitem

%%% 7
\bibitem[\protect\citeauthoryear{Wang et~al.}{2013}]{wang2013minimizing}
\begin{bchapter}
\bauthor{\bsnm{Wang}, \binits{S.}},
\bauthor{\bsnm{Ali}, \binits{S.}},
\bauthor{\bsnm{Gotlieb}, \binits{A.}}:
\bctitle{Minimizing test suites in software product lines using weight-based
  genetic algorithms}.
In: \bbtitle{Proceedings of the 15th Annual Conference on Genetic and
  Evolutionary Computation},
pp. \bfpage{1493}--\blpage{1500}
(\byear{2013})
\end{bchapter}
\endbibitem

%%% 8
\bibitem[\protect\citeauthoryear{Panichella
  et~al.}{2014}]{panichella2014improving}
\begin{barticle}
\bauthor{\bsnm{Panichella}, \binits{A.}},
\bauthor{\bsnm{Oliveto}, \binits{R.}},
\bauthor{\bsnm{Di~Penta}, \binits{M.}},
\bauthor{\bsnm{De~Lucia}, \binits{A.}}:
\batitle{Improving multi-objective test case selection by injecting diversity
  in genetic algorithms}.
\bjtitle{IEEE Transactions on Software Engineering}
\bvolume{41}(\bissue{4}),
\bfpage{358}--\blpage{383}
(\byear{2014})
\end{barticle}
\endbibitem

%%% 9
\bibitem[\protect\citeauthoryear{Arrieta et~al.}{2019}]{arrieta2019pareto}
\begin{barticle}
\bauthor{\bsnm{Arrieta}, \binits{A.}},
\bauthor{\bsnm{Wang}, \binits{S.}},
\bauthor{\bsnm{Markiegi}, \binits{U.}},
\bauthor{\bsnm{Arruabarrena}, \binits{A.}},
\bauthor{\bsnm{Etxeberria}, \binits{L.}},
\bauthor{\bsnm{Sagardui}, \binits{G.}}:
\batitle{Pareto efficient multi-objective black-box test case selection for
  simulation-based testing}.
\bjtitle{Information and Software Technology}
\bvolume{114},
\bfpage{137}--\blpage{154}
(\byear{2019})
\end{barticle}
\endbibitem

%%% 10
\bibitem[\protect\citeauthoryear{Xue and Li}{2020}]{xue2020multi}
\begin{barticle}
\bauthor{\bsnm{Xue}, \binits{Y.}},
\bauthor{\bsnm{Li}, \binits{Y.-F.}}:
\batitle{Multi-objective integer programming approaches for solving the
  multi-criteria test-suite minimization problem: Towards sound and complete
  solutions of a particular search-based software-engineering problem}.
\bjtitle{ACM Transactions on Software Engineering and Methodology (TOSEM)}
\bvolume{29}(\bissue{3}),
\bfpage{1}--\blpage{50}
(\byear{2020})
\end{barticle}
\endbibitem

%%% 11
\bibitem[\protect\citeauthoryear{Li et~al.}{2007}]{li2007search}
\begin{barticle}
\bauthor{\bsnm{Li}, \binits{Z.}},
\bauthor{\bsnm{Harman}, \binits{M.}},
\bauthor{\bsnm{Hierons}, \binits{R.M.}}:
\batitle{Search algorithms for regression test case prioritization}.
\bjtitle{IEEE Transactions on software engineering}
\bvolume{33}(\bissue{4}),
\bfpage{225}--\blpage{237}
(\byear{2007})
\end{barticle}
\endbibitem

%%% 12
\bibitem[\protect\citeauthoryear{Assun{\c{c}}{\~a}o
  et~al.}{2014}]{assunccao2014multi}
\begin{barticle}
\bauthor{\bsnm{Assun{\c{c}}{\~a}o}, \binits{W.K.G.}},
\bauthor{\bsnm{Colanzi}, \binits{T.E.}},
\bauthor{\bsnm{Vergilio}, \binits{S.R.}},
\bauthor{\bsnm{Pozo}, \binits{A.}}:
\batitle{A multi-objective optimization approach for the integration and test
  order problem}.
\bjtitle{Information Sciences}
\bvolume{267},
\bfpage{119}--\blpage{139}
(\byear{2014})
\end{barticle}
\endbibitem

%%% 13
\bibitem[\protect\citeauthoryear{Epitropakis
  et~al.}{2015}]{epitropakis2015empirical}
\begin{bchapter}
\bauthor{\bsnm{Epitropakis}, \binits{M.G.}},
\bauthor{\bsnm{Yoo}, \binits{S.}},
\bauthor{\bsnm{Harman}, \binits{M.}},
\bauthor{\bsnm{Burke}, \binits{E.K.}}:
\bctitle{Empirical evaluation of pareto efficient multi-objective regression
  test case prioritisation}.
In: \bbtitle{Proceedings of the 2015 International Symposium on Software
  Testing and Analysis},
pp. \bfpage{234}--\blpage{245}
(\byear{2015})
\end{bchapter}
\endbibitem

%%% 14
\bibitem[\protect\citeauthoryear{Di~Nucci et~al.}{2018}]{di2018test}
\begin{barticle}
\bauthor{\bsnm{Di~Nucci}, \binits{D.}},
\bauthor{\bsnm{Panichella}, \binits{A.}},
\bauthor{\bsnm{Zaidman}, \binits{A.}},
\bauthor{\bsnm{De~Lucia}, \binits{A.}}:
\batitle{A test case prioritization genetic algorithm guided by the hypervolume
  indicator}.
\bjtitle{IEEE Transactions on Software Engineering}
\bvolume{46}(\bissue{6}),
\bfpage{674}--\blpage{696}
(\byear{2018})
\end{barticle}
\endbibitem

%%% 15
\bibitem[\protect\citeauthoryear{Engstr{\"o}m
  et~al.}{2010}]{engstrom2010systematic}
\begin{barticle}
\bauthor{\bsnm{Engstr{\"o}m}, \binits{E.}},
\bauthor{\bsnm{Runeson}, \binits{P.}},
\bauthor{\bsnm{Skoglund}, \binits{M.}}:
\batitle{A systematic review on regression test selection techniques}.
\bjtitle{Information and Software Technology}
\bvolume{52}(\bissue{1}),
\bfpage{14}--\blpage{30}
(\byear{2010})
\end{barticle}
\endbibitem

%%% 16
\bibitem[\protect\citeauthoryear{Hoare and Milner}{2005}]{hoare2005grand}
\begin{barticle}
\bauthor{\bsnm{Hoare}, \binits{T.}},
\bauthor{\bsnm{Milner}, \binits{R.}}:
\batitle{Grand challenges for computing research}.
\bjtitle{The Computer Journal}
\bvolume{48}(\bissue{1}),
\bfpage{49}--\blpage{52}
(\byear{2005})
\end{barticle}
\endbibitem

%%% 17
\bibitem[\protect\citeauthoryear{Knight}{2018}]{knight2018serious}
\begin{barticle}
\bauthor{\bsnm{Knight}, \binits{W.}}:
\batitle{Serious quantum computers are finally here. what are we going to do
  with them}.
\bjtitle{MIT Technology Review. Retrieved on October}
\bvolume{30},
\bfpage{2018}
(\byear{2018})
\end{barticle}
\endbibitem

%%% 18
\bibitem[\protect\citeauthoryear{Wang et~al.}{2023}]{bootqa}
\begin{botherref}
\oauthor{\bsnm{Wang}, \binits{X.}},
\oauthor{\bsnm{Muqeet}, \binits{A.}},
\oauthor{\bsnm{Yue}, \binits{T.}},
\oauthor{\bsnm{Ali}, \binits{S.}},
\oauthor{\bsnm{Arcaini}, \binits{P.}}:
Test case minimization with quantum annealers.
arXiv:2308.05505 [cs.SE]
(2023)
\end{botherref}
\endbibitem

%%% 19
\bibitem[\protect\citeauthoryear{Wang et~al.}{2024}]{qaoa_tco}
\begin{botherref}
\oauthor{\bsnm{Wang}, \binits{X.}},
\oauthor{\bsnm{Ali}, \binits{S.}},
\oauthor{\bsnm{Yue}, \binits{T.}},
\oauthor{\bsnm{Arcaini}, \binits{P.}}:
Quantum approximate optimization algorithm for test case optimization.
IEEE Transactions on Software Engineering
(2024)
\end{botherref}
\endbibitem

%%% 20
\bibitem[\protect\citeauthoryear{D-Wave}{Accessed on: 14-11-2024}]{dwave_env}
\begin{botherref}
\oauthor{\bsnm{D-Wave}}:
D-wave environment.
[Online] \url{https://www.dwavesys.com}
(Accessed on: 14-11-2024)
\end{botherref}
\endbibitem

%%% 21
\bibitem[\protect\citeauthoryear{Farhi et~al.}{2014}]{farhi2014quantum}
\begin{botherref}
\oauthor{\bsnm{Farhi}, \binits{E.}},
\oauthor{\bsnm{Goldstone}, \binits{J.}},
\oauthor{\bsnm{Gutmann}, \binits{S.}}:
A quantum approximate optimization algorithm.
arXiv preprint arXiv:1411.4028
(2014)
\end{botherref}
\endbibitem

%%% 22
\bibitem[\protect\citeauthoryear{e~B.~K.~Chakrabart}{2008}]{Das2008Colloquium}
\begin{botherref}
\oauthor{\bsnm{B.~K.~Chakrabart}, \binits{A.D.}}:
Colloquium: Quantum annealing e analog quantum computation.
Reviews of Modern Physics, vol. 80, no. 3, p. 1061
(2008)
\end{botherref}
\endbibitem

%%% 23
\bibitem[\protect\citeauthoryear{Kadowaki and
  Nishimori}{1998}]{kadowaki1998quantum}
\begin{barticle}
\bauthor{\bsnm{Kadowaki}, \binits{T.}},
\bauthor{\bsnm{Nishimori}, \binits{H.}}:
\batitle{Quantum annealing in the transverse ising model}.
\bjtitle{Physical Review E}
\bvolume{58}(\bissue{5}),
\bfpage{5355}
(\byear{1998})
\end{barticle}
\endbibitem

%%% 24
\bibitem[\protect\citeauthoryear{Suzuki}{2009}]{suzuki2009comparison}
\begin{bchapter}
\bauthor{\bsnm{Suzuki}, \binits{S.}}:
\bctitle{A comparison of classical and quantum annealing dynamics}.
In: \bbtitle{Journal of Physics: Conference Series},
vol. \bseriesno{143},
p. \bfpage{012002}
(\byear{2009}).
\bcomment{IOP Publishing}
\end{bchapter}
\endbibitem

%%% 25
\bibitem[\protect\citeauthoryear{Rothermel and
  Harrold}{1996}]{rothermel1996analyzing}
\begin{barticle}
\bauthor{\bsnm{Rothermel}, \binits{G.}},
\bauthor{\bsnm{Harrold}, \binits{M.J.}}:
\batitle{Analyzing regression test selection techniques}.
\bjtitle{IEEE Transactions on software engineering}
\bvolume{22}(\bissue{8}),
\bfpage{529}--\blpage{551}
(\byear{1996})
\end{barticle}
\endbibitem

%%% 26
\bibitem[\protect\citeauthoryear{Fischer}{1977}]{fischer1977test}
\begin{botherref}
\oauthor{\bsnm{Fischer}, \binits{K.F.}}:
A test case selection method for the validation of software maintenance
  modifications
(1977)
\end{botherref}
\endbibitem

%%% 27
\bibitem[\protect\citeauthoryear{Yau and Kishimoto}{1987}]{yau1987method}
\begin{bchapter}
\bauthor{\bsnm{Yau}, \binits{S.S.}},
\bauthor{\bsnm{Kishimoto}, \binits{Z.}}:
\bctitle{Method for revalidating modified programs in the maintenance phase.}
In: \bbtitle{Proceedings-IEEE Computer Society's International Computer
  Software \& Applications Conference},
pp. \bfpage{272}--\blpage{277}
(\byear{1987}).
\bcomment{IEEE}
\end{bchapter}
\endbibitem

%%% 28
\bibitem[\protect\citeauthoryear{Rothermel and
  Harrold}{1993}]{rothermel1993safe}
\begin{bchapter}
\bauthor{\bsnm{Rothermel}, \binits{G.}},
\bauthor{\bsnm{Harrold}, \binits{M.J.}}:
\bctitle{A safe, efficient algorithm for regression test selection}.
In: \bbtitle{1993 Conference on Software Maintenance},
pp. \bfpage{358}--\blpage{367}
(\byear{1993}).
\bcomment{IEEE}
\end{bchapter}
\endbibitem

%%% 29
\bibitem[\protect\citeauthoryear{Bates and
  Horwitz}{1993}]{bates1993incremental}
\begin{bchapter}
\bauthor{\bsnm{Bates}, \binits{S.}},
\bauthor{\bsnm{Horwitz}, \binits{S.}}:
\bctitle{Incremental program testing using program dependence graphs}.
In: \bbtitle{Proceedings of the 20th ACM SIGPLAN-SIGACT Symposium on Principles
  of Programming Languages},
pp. \bfpage{384}--\blpage{396}
(\byear{1993})
\end{bchapter}
\endbibitem

%%% 30
\bibitem[\protect\citeauthoryear{Grover}{1996}]{grover1996fast}
\begin{bchapter}
\bauthor{\bsnm{Grover}, \binits{L.K.}}:
\bctitle{A fast quantum mechanical algorithm for database search}.
In: \bbtitle{Proceedings of the Twenty-eighth Annual ACM Symposium on Theory of
  Computing},
pp. \bfpage{212}--\blpage{219}
(\byear{1996})
\end{bchapter}
\endbibitem

%%% 31
\bibitem[\protect\citeauthoryear{Born and Fock}{1928}]{at1}
\begin{botherref}
\oauthor{\bsnm{Born}, \binits{M.}},
\oauthor{\bsnm{Fock}, \binits{V.}}:
Beweis des adiabatensatzes.
Zeitschrift f¨ ur Physik, no. 6
(1928)
\end{botherref}
\endbibitem

%%% 32
\bibitem[\protect\citeauthoryear{Morita and Nishimori}{2008}]{at2}
\begin{botherref}
\oauthor{\bsnm{Morita}, \binits{S.}},
\oauthor{\bsnm{Nishimori}, \binits{H.}}:
Mathematical foundation of quantum annealing.
Journal of Mathematical Physics, vol. 49, no. 12
(2008)
\end{botherref}
\endbibitem

%%% 33
\bibitem[\protect\citeauthoryear{Catherine and Pau}{Accessed on:
  14-11-2024}]{adv_sys}
\begin{botherref}
\oauthor{\bsnm{Catherine}, \binits{M.}},
\oauthor{\bsnm{Pau}, \binits{F.}}:
The d-wave advantage system: An overview.
D-Wave, Tech. Rep., 2022. [Online]
  \url{https://www.dwavesys.com/media/s3qbjp3s/14-1049a-a_the_d-wave_advantage_system_an_overview.pdf}
(Accessed on: 14-11-2024)
\end{botherref}
\endbibitem

%%% 34
\bibitem[\protect\citeauthoryear{McGeoch et~al.}{Accessed on:
  14-11-2024}]{tech_rep}
\begin{botherref}
\oauthor{\bsnm{McGeoch}, \binits{C.}},
\oauthor{\bsnm{Farr\'{e}}, \binits{P.}},
\oauthor{\bsnm{Bernoudy}, \binits{W.}}:
D-wave hybrid solver service + advantage: Technology update.
[Online]
  \url{https://www.dwavesys.com/media/m2xbmlhs/14-1048a-a_d-wave_hybrid_solver_service_plus_advantage_technology_update.pdf}
(Accessed on: 14-11-2024)
\end{botherref}
\endbibitem

%%% 35
\bibitem[\protect\citeauthoryear{Serrano et~al.}{2023}]{marisma_qa}
\begin{botherref}
\oauthor{\bsnm{Serrano}, \binits{M.A.}},
\oauthor{\bsnm{S{\'a}nchez}, \binits{L.E.}},
\oauthor{\bsnm{Santos-Olmo}, \binits{A.}},
\oauthor{\bsnm{Garc{\'\i}a-Rosado}, \binits{D.}},
\oauthor{\bsnm{Blanco}, \binits{C.}},
\oauthor{\bsnm{Barletta}, \binits{V.S.}},
\oauthor{\bsnm{Caivano}, \binits{D.}},
\oauthor{\bsnm{Fern{\'a}ndez-Medina}, \binits{E.}}:
Minimizing incident response time in real-world scenarios using quantum
  computing.
Software Quality Journal,
1--30
(2023)
\end{botherref}
\endbibitem

%%% 36
\bibitem[\protect\citeauthoryear{Fred~Glover}{2022}]{qubo_tutorial}
\begin{botherref}
\oauthor{\bsnm{Fred~Glover}, \binits{R.H.e.Y.D.} \bsuffix{Gary~Kochenberger}}:
Quantum bridge analytics i: A tutorial on formulating and using qubo models.
Ann Oper Res 314, 141–183
(2022)
\end{botherref}
\endbibitem

%%% 37
\bibitem[\protect\citeauthoryear{Steuer}{1986}]{multi_opt}
\begin{botherref}
\oauthor{\bsnm{Steuer}, \binits{R.E.}}:
Multi-criteria optimization: Theory, computation, and application.
John Wiley, New York
(1986)
\end{botherref}
\endbibitem

%%% 38
\bibitem[\protect\citeauthoryear{Ayodele}{2022}]{penalty_weights}
\begin{botherref}
\oauthor{\bsnm{Ayodele}, \binits{M.}}:
Penalty weights in qubo formulations: Permutation problems
(2022)
\end{botherref}
\endbibitem

%%% 39
\bibitem[\protect\citeauthoryear{SIR}{Accessed on: 14-11-2024}]{sir_rep}
\begin{botherref}
\oauthor{\bsnm{SIR}}:
Software-artifact infrastructure repository.
[Online] \url{https://sir.csc.ncsu.edu/portal/index.php}
(Accessed on: 14-11-2024)
\end{botherref}
\endbibitem

%%% 40
\bibitem[\protect\citeauthoryear{Trovato}{Accessed on: 14-11-2024}]{appendix}
\begin{botherref}
\oauthor{\bsnm{Trovato}, \binits{A.}}:
Appendix.
[Online] \url{https://github.com/AntonioTrovato/SelectQA}
(Accessed on: 14-11-2024)
\end{botherref}
\endbibitem

%%% 41
\bibitem[\protect\citeauthoryear{Conover}{1998}]{t_test}
\begin{botherref}
\oauthor{\bsnm{Conover}, \binits{W.J.}}:
Practical nonparametric statistics.
3rd ed. New York, NY, USA: Wiley
(1998)
\end{botherref}
\endbibitem

%%% 42
\bibitem[\protect\citeauthoryear{Vargha and Delaney}{2000}]{a12}
\begin{botherref}
\oauthor{\bsnm{Vargha}, \binits{A.}},
\oauthor{\bsnm{Delaney}, \binits{H.D.}}:
A critique and improvement of the cl common language effect size statistics of
  mcgraw and wong.
J. Educ. and Behav. Stat. 25(2), 101–132
(2000)
\end{botherref}
\endbibitem

%%% 43
\bibitem[\protect\citeauthoryear{Robotics}{Accessed on:
  14-11-2024}]{paintcontrol_ds}
\begin{botherref}
\oauthor{\bsnm{Robotics}, \binits{A.}}:
Industrial robot supplier and manufacturer.
[Online] \url{http://new.abb.com/products/robotics}
(Accessed on: 14-11-2024)
\end{botherref}
\endbibitem

%%% 44
\bibitem[\protect\citeauthoryear{Google}{Accessed on: 14-11-2024}]{gsdtsr}
\begin{botherref}
\oauthor{\bsnm{Google}}:
Google code archive.
[Online]
  \url{https://code.google.com/archive/p/google-shared-dataset-of-test-suite-results/wikis/DataFields.wiki}
(Accessed on: 14-11-2024)
\end{botherref}
\endbibitem

%%% 45
\bibitem[\protect\citeauthoryear{Muqeet}{Accessed on: 14-11-2024}]{bootqa_rep}
\begin{botherref}
\oauthor{\bsnm{Muqeet}, \binits{A.}}:
Bootqa.
[Online] \url{https://github.com/AsmarMuqeet/BootQA}
(Accessed on: 14-11-2024)
\end{botherref}
\endbibitem

\end{thebibliography}
%% if required, the content of .bbl file can be included here once bbl is generated
%%\input sn-article.bbl

\end{document}